\documentclass[a4paper,english,10pt,twoside]{article}
\usepackage[english]{babel}
\usepackage[latin1]{inputenc}
\usepackage{amsmath}
\usepackage{graphicx}

\usepackage{fancyhdr}
\fancypagestyle{plain}{%
  \fancyhf{}%
  \fancyhead[C]{\small\it{Twenty years of giant exoplanets - Proceedings of the Haute Provence Observatory Colloquium, 5-9 October 2015\\ Edited by I. Boisse, O. Demangeon, F. Bouchy \& L. Arnold} }
  \fancyfoot[C]{\normalsize \thepage}%
}

\newcommand{\nc}   {\newcommand}
\nc{\dsfrac}[2]{{\displaystyle\frac{#1}{#2}}}

\newcommand{\Jcal}  {{\EuScript J}}

\nc{\ergs}  {\mathrm{erg}/\mathrm{gm\;s}}
\nc{\ergc}  {{\rm erg}~{\rm cm}^{-3}}
\nc{\ergcs} {{\rm erg}~{\rm cm}^{-2}~{\rm s}^{-1}}
\nc{\ms}   {{\rm m}~{\rm s}^{-1}}

\nc{\Eem}   {\widetilde{E}_\mathrm{e}}
\nc{\Eexm}  {\widetilde{E}_\mathrm{ex}}
\nc{\Eim}   {\widetilde{E}_\mathrm{I}}
\nc{\ER}    {E_\mathrm{R}}
\nc{\FL}    {F_\mathrm{Ly}}
\nc{\FR}    {F_\mathrm{R}}
\nc{\FRa}   {F_\mathrm{R,1}}
\nc{\FFR}   {\mathbf{F}_\mathrm{R}}
\nc{\Fe}    {F_\mathrm{e}}
\nc{\FFe}   {\mathbf{F}_\mathrm{e}}
\nc{\Fsat}  {q_\mathrm{sat}}
\nc{\Ke}    {\EuScript{K}_\mathrm{e}}
\nc{\mfp}   {l_\mathrm{e}}
\nc{\mrc}   {\mathrm{c}}
\nc{\mrr}   {\mathrm{r}}
\nc{\nH}    {n_\mathrm{H}}
\nc{\nel}   {n_\mathrm{e}}
\nc{\Pe}    {P_\mathrm{e}}
\nc{\Pg}    {P_\mathrm{g}}
\nc{\PR}    {P_\mathrm{R}}
\nc{\Qelc}  {Q_\mathrm{elc}}
\nc{\Qinc}  {Q_\mathrm{inc}}
\nc{\gyr}   {r_\mathrm{B}}
\nc{\Ta}    {T_a}

\nc{\divFe} {\nabla\cdot\FFe}
\nc{\divFR} {\nabla\cdot\FFR}
\nc{\FLj}   {F_\mathrm{Ly,\Jcal}}
\nc{\FRj}   {F_\mathrm{R,\Jcal}}

\nc{\aap}   {A\&A}
\nc{\aaps}   {A\&AS}
\nc{\apj}   {ApJ}
\nc{\mnras}   {MNRAS}
\nc{\aapr} {A\&ARv}
\nc{\pasp} {PASP}
\usepackage{txfonts}
\usepackage[squaren,cdot]{SIunits}
\usepackage{tabularx}
\usepackage{lscape}
\usepackage[flushleft]{threeparttable}
\usepackage{array,longtable}
\usepackage[breaklinks=true]{hyperref} 
\usepackage{natbib,twoopt}
\bibpunct{(}{)}{;}{a}{}{,} 
\makeatletter
\newcommandtwoopt{\citeads}[3][][]{\href{http://adsabs.harvard.edu/abs/#3}%
{\def\hyper@linkstart##1##2{}%
\let\hyper@linkend\@empty\citealp[#1][#2]{#3}}}
\newcommandtwoopt{\citepads}[3][][]{\href{http://adsabs.harvard.edu/abs/#3}%
{\def\hyper@linkstart##1##2{}%
\let\hyper@linkend\@empty\citep[#1][#2]{#3}}}
\newcommandtwoopt{\citetads}[3][][]{\href{http://adsabs.harvard.edu/abs/#3}%
{\def\hyper@linkstart##1##2{}%
\let\hyper@linkend\@empty\citet[#1][#2]{#3}}}
\newcommandtwoopt{\citeyearads}[3][][]%
{\href{http://adsabs.harvard.edu/abs/#3}
{\def\hyper@linkstart##1##2{}%
\let\hyper@linkend\@empty\citeyear[#1][#2]{#3}}}
\makeatother

\makeatletter

\def\@maketitle{%
  \vskip 2em%
  \begin{center}%
  \let \footnote \thanks
    {\LARGE\textbf \@title \par}%
    \vskip 1.5em%
    {\normalsize
      \lineskip .5em%
      \begin{tabular}[t]{c}%
        \@author
      \end{tabular}\par}%
    \vskip 1em%
    {\normalsize \@date}%
  \end{center}%
  \par
  \vskip 1.5em}
\makeatother

\newcommand{\affil}[1]{\small{\hskip-0.55cm #1}}

\begin{document}

\setcounter{page}{1}  

\title{Eight years of accurate photometric follow-up of transiting giant exoplanets}
\author{ L. Mancini$^{1}$,  J. Southworth$^{2}$}
\date{} 
\maketitle
\affil{  $^1$Max Planck Institute for Astronomy, K\"{o}nigstuhl 17, 69117 -- Heidelberg, Germany (\texttt{mancini@mpia.de})\\
          $^2$Astrophysics Group, Keele University, Staffordshire, ST5 5BG, UK (\texttt{astro.js@keele.ac.uk})
          }

\vskip1cm

\begin{abstract}
Since 2008 we have run an observational program to accurately measure the characteristics of known exoplanet systems hosting close-in transiting giant planets, i.e.\ hot Jupiters.  Our study is based on high-quality photometric follow-up observations of transit events with an array of medium-class telescopes, which are located in both the northern and the southern hemispheres. A high photometric precision is achieved through the {\it telescope-defocussing} technique. The data are then reduced and analysed in a homogeneous way for estimating the orbital and physical parameters of both the planets and their parent stars. We also make use of multi-band imaging cameras for probing planetary atmospheres via the transmission-photometry technique. In some cases we adopt a two-site observational strategy for collecting simultaneous light curves of individual transits, which is the only completely reliable method for truly distinguishing a real astrophysical signal from systematic noise. In this contribution we review the main results of our program.
\end{abstract}

\section{Introduction}

Transiting extrasolar planets (TEPs) are the most important and interesting planets to study. The particular orbital configuration of these planetary systems, with respect to an Earth-based observer, enables measurement of their main physical properties, including the planet's mass, radius, density, surface gravity and temperature, which are of huge importance for finding Earth twins in {\it habitable} zones around normal stars. Specific science cases for this work include high-precision measurements of the properties of planetary systems, transmission photometry and spectroscopy to study the atmospheres of giant planets, and transit timing work to study the dynamical properties of planetary systems.

In August 2008, we began a long-term observational program for measuring the physical properties of know TEP systems via accurate photometric monitoring of transit events. Our project was conceived considering: ($i$) the poor quality of the photometric data on which many TEP discoveries were based; and ($ii$) the necessity of analysing the data in a homogeneous way, so that the properties of different exoplanets can be reasonably comparable in a global picture. Establishing such a homogeneous and trustworthy dataset is fundamental for theoretical studies of how exoplanets form and evolve.

The scope of our project has inexorably grown alongside the discovery rate of suitable planets for analysis. The time-critical nature of transit observations naturally encourages a {\it survey-style} project whereby large numbers of transits are scheduled for observation in order to overcome the difficulties of scheduling and losses due to weather and technical issues. Once a sufficient number of transits have been observed for a given planet, these can then be used to characterise the planetary system in detail. The need for a large amount of observing time is best met by using ground-based medium-class telescopes, which are sufficient for the project and more readily available than larger facilities. These telescopes are best suited to the study of close-in giant planets, usually termed ``hot Jupiters'', orbiting bright stars ($V\lesssim 14$\,mag). For these TEPs, the transits are deep (typically 1--2\%) and frequent, and a high photometric precision can be obtained using 1--2\,m telescopes.

\subsection{Hot Jupiters}

Hot Jupiters arguably represent the first class of exoplanets found, and 51\,Peg\,b is their prototype. They are giant gas planets\footnote{According to the definition given by \citet{hatzes:2015}, giant planets cover the mass range $0.3-60\,M_{\rm Jup}$.} with tight ($\sim 0.01-0.05$\,au) and short-period ($\sim 1-10$\,days) orbits around their parent stars. They are strongly irradiated by their host stars, resulting in high equilibrium temperatures (e.g.\ 2750\,K for Kepler-13 and 2710\,K for WASP-33). Although it appears that they are very uncommon with respect to Neptunian and rocky planets (e.g. \citealp{fressin:2013,petigura:2013}), there are many motivations for studying them. Their relatively large mass and radius allows measurement of these quantities to much better precision than smaller planets, their spin-orbit alignment is directly accessible by observing the Rossiter-McLaughlin effect, transmission spectra can be obtained of the terminator regions of their atmospheres, and their day-side thermal emission and reflected light are measurable. It is therefore possible to investigate the properties of their atmospheres and the abundances of elements and molecules. However, after four {\it lustra} from their discovery \citep{mayor:1995}, their formation and evolution mechanisms are still unclear and under intriguing investigation and debate. In particular, it is not clear what are the physical mechanisms responsible of their migration from the snow line ($\sim 3$\,au), where they should form, down to roughly $10^{-2}$\,au from their parent stars.

\section{Observations}

The medium-class telescopes, which we utilise in our program, summarised in Table\,\ref{table_1}, are equipped with CCD cameras that have fields-of-view (FOVs) of up to several tens of arcmin, allowing the possibility to include in the scientific images a good number ($\sim 3-10$) of reference stars, which are vital for achieving high-quality differential photometry.
\begin{table}[h]
\caption{List of the telescopes used in our program}
\centering
\begin{tabular}{lccccc}
\hline
Telescope & Observatory & Aperture & Instrument & Multi-band ability & Transits \\
\hline\\[-6pt]
\multicolumn{2}{l}{\textbf{Southern hemisphere}} \\ [2pt] %
MPG 2.2\,m  & La Silla        & 2.2\,m    & GROND     & 4 bands   &  49 \\
Danish      & La Silla        & 1.54\,m   & DFOSC     & No        & 148 \\ [4pt] %
\multicolumn{2}{l}{\textbf{Northern hemisphere}} \\ [2pt] %
INT         & La Palma        & 2.5\,m    & WFC       & No        &  32 \\
CAHA 2.2\,m & Calar Alto      & 2.2\,m    & BUSCA     & 4 bands   &  33 \\
Cassini     & Loiano          & 1.52\,m   & BFOSC     & No        &  72 \\
Zeiss       & Calar Alto      & 1.23\,m   & DLR-MKIII & No        & 151 \\
\hline\hline
\end{tabular}
\label{table_1}
\end{table}
We perform photometric observations of planetary transits through broad-band filters, generally Cousins/Bessell $R$ and $I$ or Sloan/Gunn $r$ and $i$ (according to the magnitude and colour of the parent stars). This choice is dictated by several considerations: ($i$) we generally observe cool dwarf stars, which emit more radiation between 6000 and 8000\,\AA; ($ii$) limb darkening (LD) is weaker than at bluer wavelengths so the transit light curves are more box-shaped and thus the transit depth and timings of the four contact points are easier to measure; ($iii$) at these wavelengths, the photometry is less affected by extinction from Earth's atmosphere and from spot activity on the surfaces of the parent stars.

\subsection{Telescope defocussing}

\begin{figure*}[h]
  \centering
  \includegraphics[width=14cm]{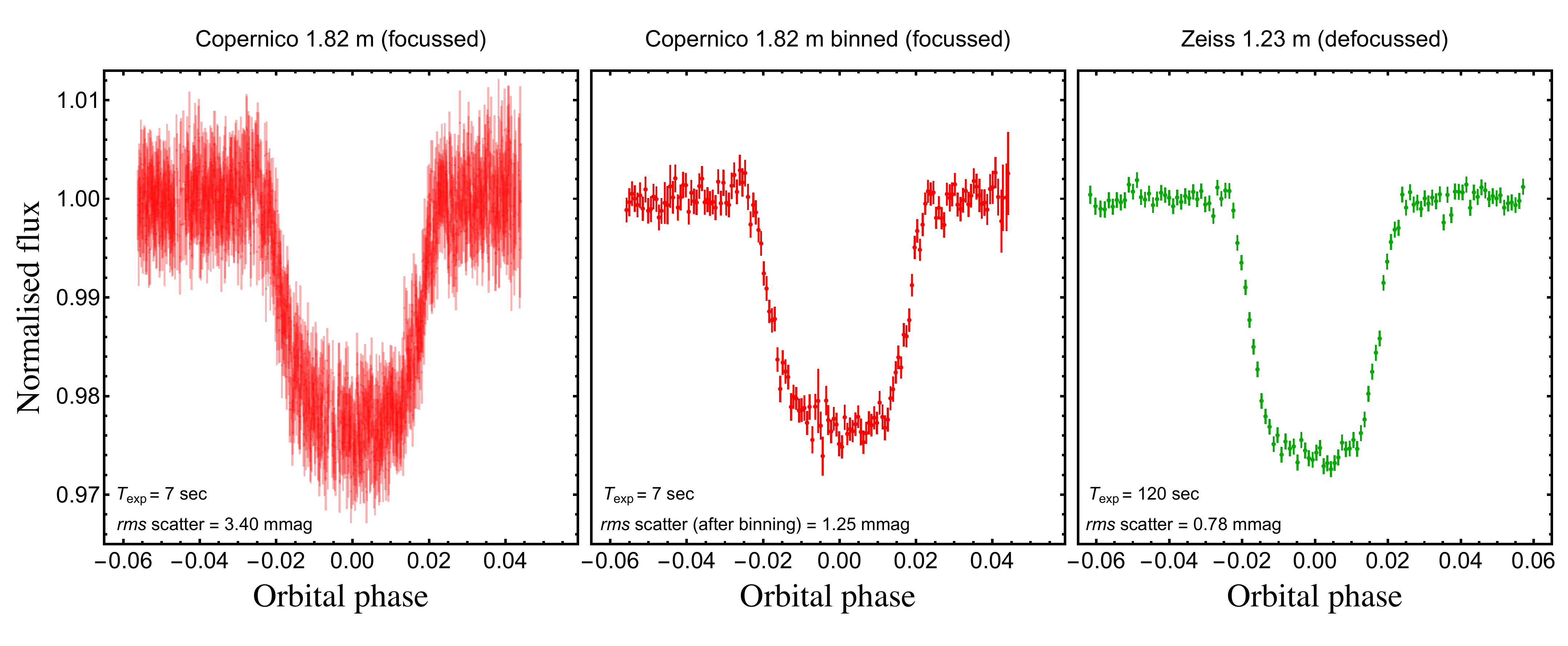}
  \caption{Example light curves of Qatar-1 from \citet{covino:2013}. {\it Left}: follow-up light curve from a 1.82\,m telescope operated in focus with an exposure time of 7\,sec. {\it Centre}: the same light curve after binning. {\it Right}: follow-up light curve from a defocussed 1.23\,m telescope with an exposure time of 120\,sec.}
  \label{figure1}
\end{figure*}

If the target is not too close to nearby stars, the observations are performed using the {\it telescope-defocusing} technique \citep{alonso:2008,southworth:2009a}, which allows us to get light curves with a higher precision than using telescopes operated in focus\footnote{It is a fairly common strategy to make observations with the telescope in focus and then bin the data points. While this method can yield light curves with quite high photometric precision, it is much more strongly affected by red noise from effects such as flat-fielding imperfections.} (see Fig.\,\ref{figure1}).
%
%
This observational method consists of defocussing the telescope so that we can use long exposure times (with a maximum of $\sim 120$\,sec to ensure we get a good sampling of the transit event) and collect more photons in each point spread function (PSF), which can now cover thousands of pixels. The immediate result is that the light from the target and the stars in the FOV show up on the CCD in the particular shape of annuli of different sizes. Collecting more photons during longer exposures greatly lowers the Poisson and scintillation noise, and systematic noise coming from flat-fielding and focus or seeing variations. We stress that there is not a general setting for the exposure time and the amount of defocussing, but they must be chosen case by case, by considering the brightness of the target and reference stars, the aperture of the telescope and the filter used. Special attention is placed in tuning the defocussing in order to avoid to work in the non-linear regime of the CCDs (we usually try to get a maximum counts per pixel around 25000--35000\,ADU for 16-bit CCD systems). Once the set-up is established, a changing of the exposure time during the monitoring of planetary-transit events (they generally last several hours) is possible if we need to compensate for variation in seeing, airmass and sky transparency, which affect the count rate of the observations during the sequence.

More rarely, the defocus can also be adjusted under particular circumstances. At several observatories, one could see the telescope going slowly from defocussed to in-focus after tracking past the meridian. This is generally caused by a gradual shifting of the primary mirror under the changing influence of gravity. This causes the observer to greatly decrease the exposure time, thus losing the benefits of the defocussing technique, and also results in PSFs of very different sizes between the first and the second part of the observations. One way to solve the problem is to take a note of the number of pixels inside the annulus of the target star at the beginning of the observations, and periodically check that this number remains constant during the sequence. If it changes (i.e.\ the size of the annulus became smaller), we increase the defocus to have again the same number of pixels. Indeed, what it is important is to keep the same amount of defocus, i.e.\ the width of the PSF for each star in the FOV should cover the same number of pixels through the full observing sequence.

\subsection{Two-telescope observations}

\begin{figure*}[h]
  \centering
  \includegraphics[width=14cm]{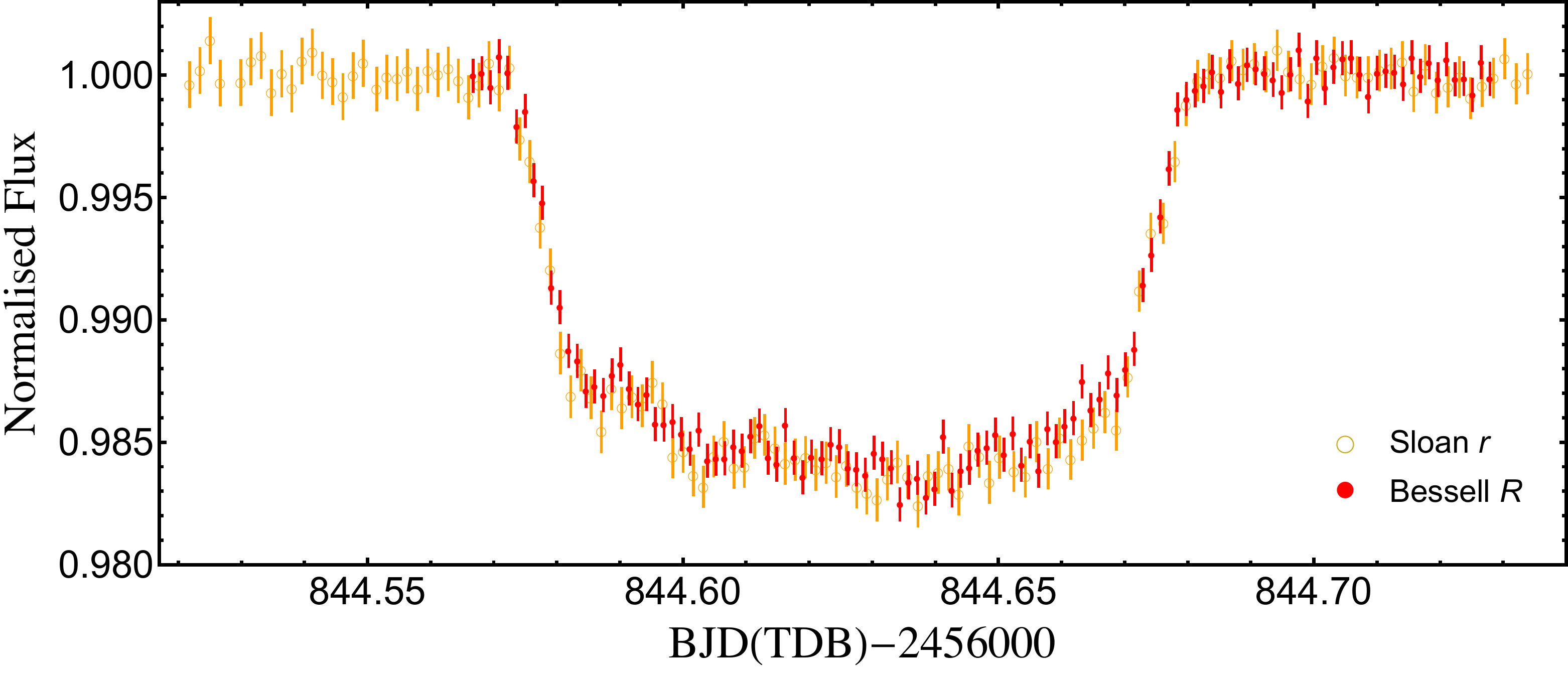}
  \caption{Simultaneous observations of a transit of WASP-103\,b taken with the MPG 2.2\,m and the Danish 1.54\,m telescopes \citep{southworth:2015a}. The same features (probably a starspot bell or granulation) are very well recorded by the two telescopes during the scanning of the stellar photosphere through the planetary transit.}
  \label{figure2}
\end{figure*}

Time series photometry of transit events can be affected by many sources of systematics, which are not always easily found and corrected, but also by astrophysical phenomena such as gravity darkening, the presence of exomoons or Trojan bodies, and the planet crossing irregularities on the stellar photosphere such a starspot or a star-spot complex. The question is how we can be completely sure that we have identified all the systematics in our data and distinguish real astrophysical signals from them? The only way to be completely sure is to observe the same transit event with two different telescopes through similar filters, preferably at different observatories. If we see the same anomaly/feature occurring at the same time on both light curves, than we can unambiguously claim that it is of astrophysical origin. We adopted this observational strategy in only a few cases (see an example in Fig.\,\ref{figure2}), as telescope scheduling, technical problems and weather conditions can all frustrate simultaneous observations.
%

\subsection{Simultaneous multi-band observations}

The radius of an exoplanet, which we measure from a transit event, can vary as a function of wavelength. Variations of its atmospheric opacity can cause the absorption of light rays coming from the parent star at specific wavelengths, allowing detection of the presence of atomic (e.g.\ Na and K) and molecular (e.g.\ H$_2$O and CO) species. This means that by photometrically observing a transit through different passbands, we can reconstruct the transmission spectra of TEPs and probe their atmospheres. This technique, also known as {\it transmission photometry}, is similar to transmission spectroscopy, with the clear disadvantage of a lower spectral resolution. On the other hand, there are many benefits to prefer photometry to spectroscopy, mainly the possibility to use ground-based telescopes with smaller apertures, the possibility to investigate TEPs orbiting faint stars, more relaxed constraints on comparison stars, and the established fact that photometry is much less affected by telluric absorption and changing airmass and sky conditions. Indeed, the most robust results of planetary-atmosphere investigation come from space observations with the HST (e.g.\ \citealp{sing:2015,nikolov:2015}). However these studies are limited by the orbital period ($\sim 90$\,min) of the spacecraft around Earth, causing a target star to be unobservable for half of each orbit. This implies that, in order to have data covering a complete transit and part of the out-of-transit light curve, observations with the HST have to be performed over multiple transits, which are generally spread over many days, during which the telescope suffers continual changes in temperature and focus. Moreover, since occulted and un-occulted starspots can cause a wavelength-dependent variations of the transit depth \citep{sing:2011}, it is also always better to obtain the observations of complete transits at multiple wavelengths simultaneously. The use of multi-imaging cameras are thus mandatory for performing transmission photometry.

\begin{figure*}[h]
  \includegraphics[width=14.5cm]{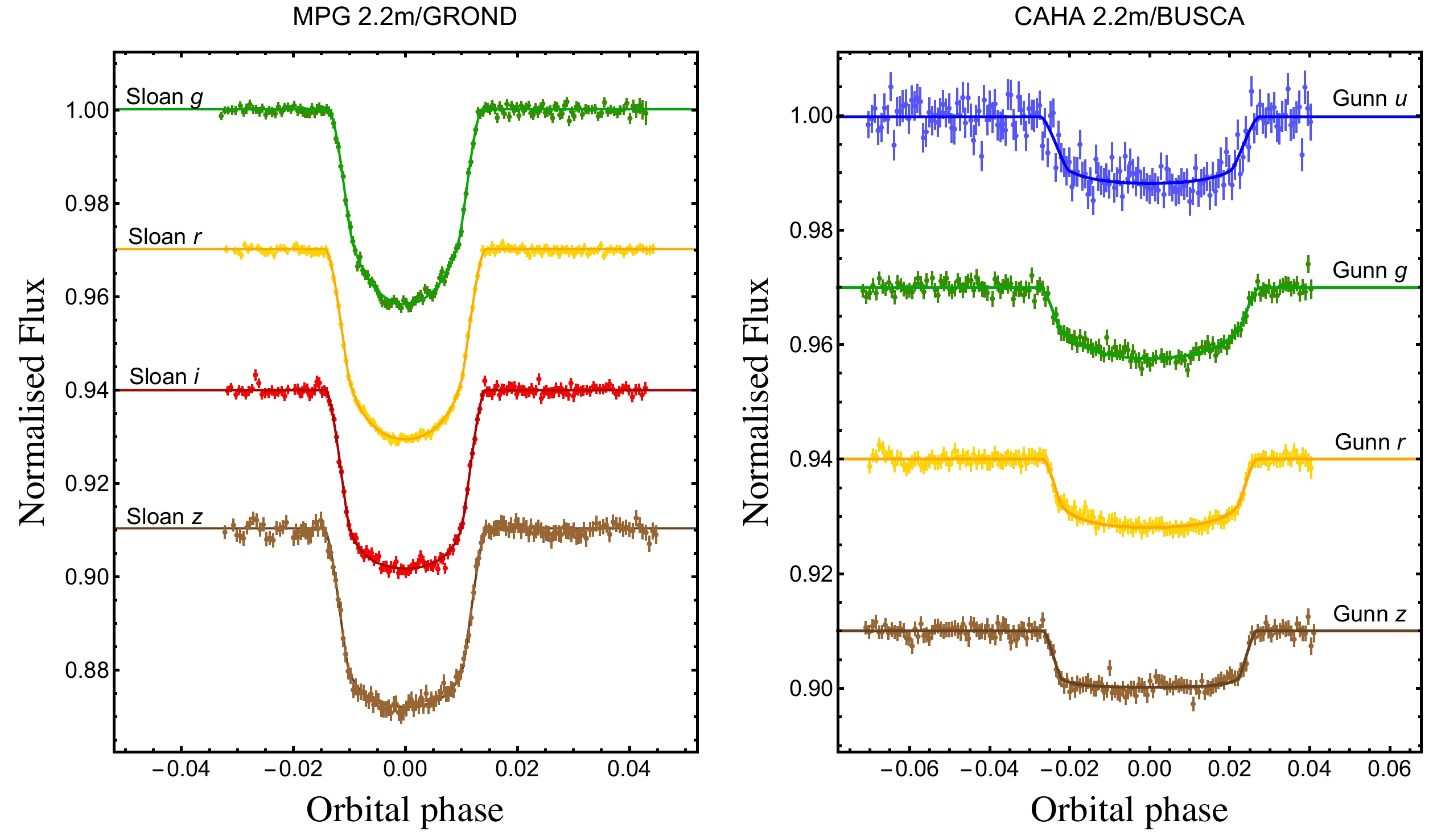}
  \caption{Examples of typical light curves obtained with multi-band imagers.  {\it Left-hand panel}: light curves of a transit of WASP-80\,b observed with GROND \citep{mancini:2014a}. {\it Right-hand panel}: light curves of a transit of HAT-P-8\,b observed with BUSCA \citep{mancini:2013a}.}
  \label{figure3}
\end{figure*}

We use two instruments that are able to perform simultaneous multi-band photometric observations (see examples in Fig.\,\ref{figure3}): ($i$) the Gamma-Ray Burst Optical and Near-Infrared Detector (GROND) instrument, mounted on the MPG 2.2\,m telescope, capable of simultaneous observations in four broad optical (similar to Sloan $g$, $r$, $i$, $z$) and three NIR ($J$, $H$, $K$) passbands \citep{greiner:2008}; ($ii$) the Bonn University Simultaneous CAmera (BUSCA), mounted on the 2.2\,m telescope at Calar Alto, able to simultaneously make photometry in four wavelength bands (from UV to visual IR), with the possibility to use different sets of filters, both broad and narrow \citep{reif:1999}. Neither instrument was designed for exoplanet work, and they therefore have operational limitations which lower the quality of the data obtainable. These limitations include needing to use the same exposure times and focus settings in all passbands, resulting in low count rates in the bluer passbands where the stars are fainter, a small field of view which limits the number of comparison stars available, and non-standard passbands which are broad or have low throughput.

\section{Light-curve analysis}

Once we have collected photometric observations of planetary transits, they are analysed in three steps following the {\it Homogeneous Studies} approach (\citealp{southworth:2012hs}, and references therein).

\begin{itemize}

\item[($i$)] We reduce the photometric data of the observed transits using {\sc defot}, an {\sc idl}\footnote{{\sc idl} is a trademark of the ITT Visual Information Solutions.}-based data reduction pipeline designed specifically for time-series photometry of defocussed images. After having calibrated the scientific images, we choose an optimal ensemble of comparison stars and perform {\it standard} aperture photometry to extract the light curves. We then remove instrumental and astrophysical trends from our light curves by fitting a straight line or a polynomial to the out-of-transit data.

\item[($ii$)] We use {\sc jktebop}\footnote{The source code of {\sc jktebop} is available at http://www.astro.keele.ac.uk/jkt/codes/jktebop.html} \citep{southworth:2012hs} for modelling the transit light curves.
This code is able to fit the orbital inclination, $i$, the transit midpoint, $T_0$, the orbital period, $P_{\rm orb}$, the limb-darkening coefficients, and the sum and ratio of the fractional radii of the star and planet, $r_{\star} + r_{\rm p}$ and $k = r_{\rm p}/r_{\star}$. Here the fractional radii are defined as $r_{\star} = R_{\star}/a$ and $r_{\rm p} = R_{\rm p}/a$, where $a$ is the orbital semi-major axis, and $R_{\star}$ and $R_{\rm p}$
are the absolute radii of the star and the planet, respectively. The uncertainties in the parameters are robustly determined by Monte Carlo simulations, bootstrapping simulations, and/or a residual-permutation algorithm.

If anomalies, which are generally caused by starspots occulted by the planet during the transit event, are present in the light curves, then we model these with the {\sc prism}+{\sc gemc} codes \citep{tregloanreed:2013,tregloanreed:2015}. In this case, the fitted parameters are the same as for {\sc jktebop} but with additional parameters to describe the starspots. Each spot is specified using the longitude and colatitude of its centre, its angular radius and its contrast, i.e.\ the ratio of the surface brightness of the star-spot to that of the surrounding photosphere.

\item[($iii$)] The full physical properties of a transiting planetary system are not directly accessible from observations alone, so the measured parameters from the light curves and spectroscopic observations must be combined with an additional constraint. For this we use the {\sc jktabsdim} code \citep[see][]{southworth:2009hs}, which can use either theoretical predictions of the properties of low-mass stars or an empirical calibration from eclipsing binary stars \citep[see][]{southworth:2010hs} as this additional constraint. Five different sets of theoretical models are implemented, allowing some idea of the systematic differences between the models to be obtained. {\sc jktabsdim} takes as input $r_{\star}$, $r_{\rm p}$, $i$ and $P_{\rm orb}$ from the light curve analysis, and the orbital eccentricity, velocity amplitude, effective temperature and metallicity from spectroscopic analysis of the host star. It returns the best-fitting properties of the star and planet, and the age of the system.

\end{itemize}

\section{Results and future perspectives}
\begin{figure*}[h]
  \centering
  \includegraphics[width=14cm]{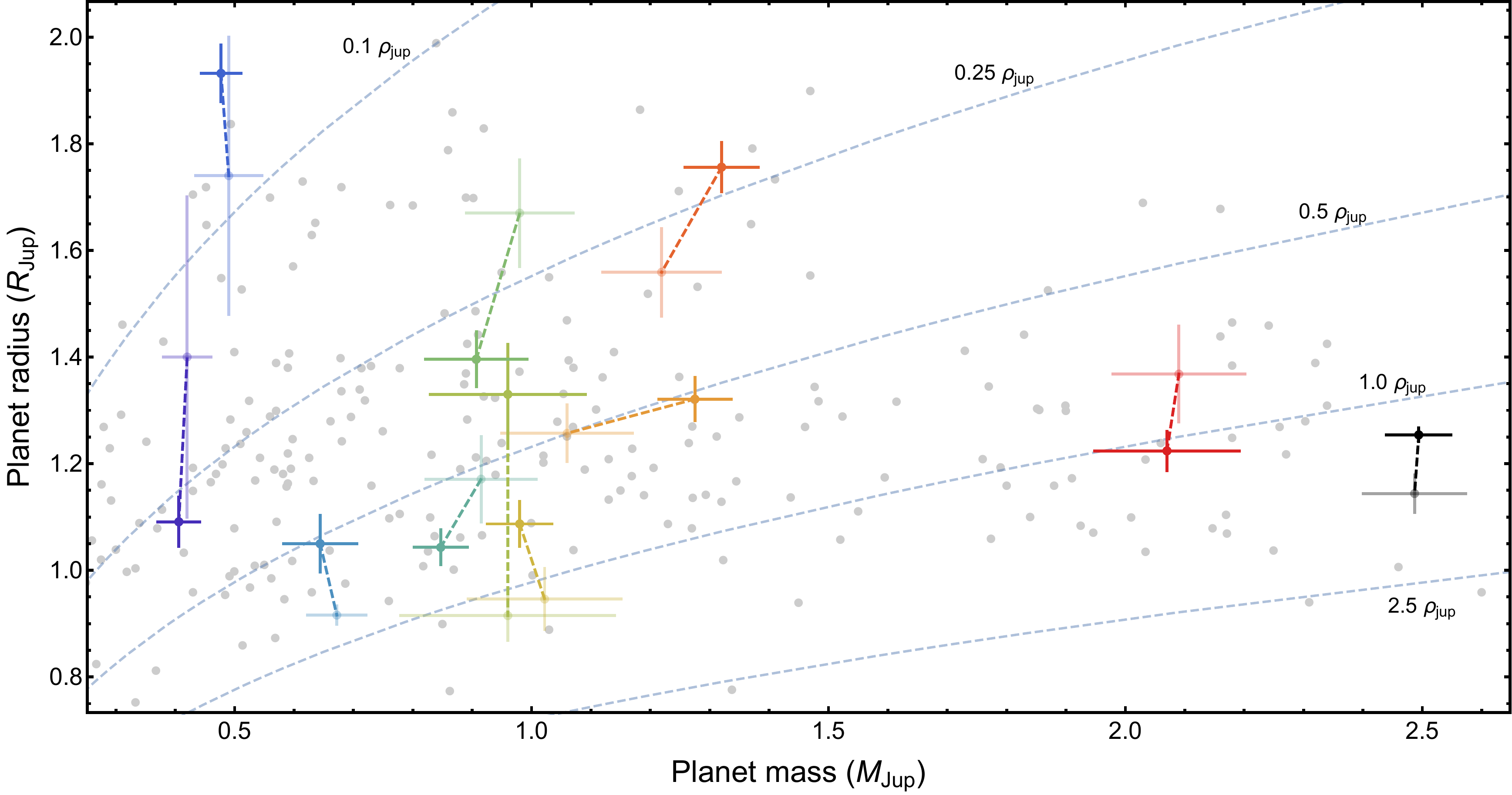}
  \caption{Masses and radii of the known transiting extrasolar planets. The diagram is zoomed on the parameter space in which most of the transiting hot-Jupiter planets have been detected. We report some emblematic cases that come from our program (dark colours), compared with the values from the discovery papers (light colours). The values of the planets represented with grey points were taken from the TEPCat catalogue (www.astro.keele.ac.uk/jkt/tepcat/; \citealp{southworth:2011tepcat}) and their errorbars have been suppressed for clarity. Dotted lines show where density is 2.5, 1.0, 0.5, 0.25 and 0.1 $\rho_{\rm Jup}$.}
  \label{figure4}
\end{figure*}

We have obtained high-quality light curves for more than 50 known TEPs and, so far, have refined the main physical parameters of 38 of them, which are summarised in Table\,\ref{Tab:results}. In most of the cases and for most of the parameters we obtained better estimations. In particular, for several TEPs, we found that their measured characteristics are significantly different to previous works based on less data. Some of these cases are highlighted in Fig.\ref{figure4}, which shows large shifts in the measured properties for some planets. The largest changes were seen for WASP-7 and WASP-16, where the measured planetary densities decreased by a factor of two or more.

Thanks to the multi-band observations performed with GROND and BUSCA, we have probed the atmosphere, at the terminator region, of 14 TEPs; they are:
HAT-P-5\,b, HAT-P-8\,b, HAT-P-23\,b, Qatar-2\,b, WASP-19\,b, WASP-36\,b, WASP-44\,b, WASP-45\,b, WASP-46\,b, WASP-48\,b, WASP-57\,b, WASP-67\,b, WASP-80\,b, WASP-103\,b. In several cases we have founded larger planet radius at bluer optical wavelengths. In particular, for the hot Jupiters WASP-36\,b \citep{mancini:2016} and WASP-103\,b \citep{southworth:2015a} we have determined a large variation between blue ($g$-band) and red ($z$-band) optical wavelengths of roughly 10 atmospheric pressure scaleheight to a confidence level of more than 5 and 7 $\sigma$. In both the cases, this variation is too large to be attributable to Rayleigh scattering in the planetary atmosphere, and should be caused by the presence of strong absorber species at bluer wavelengths.

The accurate determination of the physical properties of TEPs systems via ground-based photometric follow-up observations will continue until the Transiting Exoplanet Survey Satellite (TESS) will start its operations in late 2017. Indeed, this part of our project will superseded by the higher photometric precision of the TESS telescope. However, in the next future, there will be no space missions completely dedicated to the study of planet atmospheres. This means that for still many years, ground-based multi-band photometric observations can play an important role for investigating the properties and chemical composition of the TEPs' atmospheres, especially for those orbiting around faint stars, strongly cooperating with ground-based spectrometers and observations performed by the HST, the Spitzer and, soon, the JWST spacecrafts. It is then important, as suggested by \citet{southworth:2015b}, to have a new multi-band camera, specifically designed for transit observations. This should have a sufficient number of passbands for covering the entire optical transmission spectrum of a exoplanet and the possibility to set a different exposure time for each filter, for guaranteeing enough S/N in each band.

\begin{landscape}
\begin{center}
{\tiny
\begin{longtable}{ll|cccc|cccc|cc}
\caption[Revised parameters of the TEP systems that we have studied]{Revised parameters of the TEP systems that we have studied} 
\label{Tab:results} \\
\hline
\hline 
\multicolumn{1}{c}{TEP system} & \multicolumn{1}{c|}{References} & \multicolumn{1}{c}{$M_{\star} (M_{\odot})$}  &
\multicolumn{1}{c}{$R_{\star} (R_{\odot})$} & \multicolumn{1}{c}{$\log{g}_{\star}$(cgs)} & \multicolumn{1}{c|}{$\rho_{\star} (\rho_{\odot})$} &
\multicolumn{1}{c}{$M_{\rm p}  (M_{\rm Jup})$} & \multicolumn{1}{c}{$R_{\rm p} (R_{\rm Jup})$} & \multicolumn{1}{c}{$g_{\rm p}$(m/s$^2$)} &
\multicolumn{1}{c|}{$\rho_{\rm p} (\rho_{\rm Jup})$} & \multicolumn{1}{c}{$a$\,(au)} & \multicolumn{1}{c}{$P$\,(days))} 
\\
\hline 
\endfirsthead
\multicolumn{3}{c}%
{{\bfseries \tablename\ \thetable{} -- continued from previous page}} \\
\hline
\hline 
\multicolumn{1}{c}{TEP system} & \multicolumn{1}{c}{References} & \multicolumn{1}{c}{$M_{\star} (M_{\odot})$}  &
\multicolumn{1}{c}{$R_{\star} (R_{\odot})$} & \multicolumn{1}{c}{$\log{g}_{\star}$(cgs)} & \multicolumn{1}{c}{$\rho_{\star} (\rho_{\odot})$} &
\multicolumn{1}{c}{$M_{\rm p}  (M_{\rm Jup})$} & \multicolumn{1}{c}{$R_{\rm p} (R_{\rm Jup})$} & \multicolumn{1}{c}{$g_{\rm p}$(m/s$^2$)} &
\multicolumn{1}{c}{$\rho_{\rm p} (\rho_{\rm Jup})$} & \multicolumn{1}{c}{$a$\,(au)} & \multicolumn{1}{c}{$P$\,(days))} 
\\
\hline 
\endhead
\hline \multicolumn{2}{l}{{{\bf Continued on next page}}} \\ 
\hline
\endfoot
\hline 
\endlastfoot
%
HAT-P-5 & 
\begin{tabular}{l}
\citet{southworth:2012b} \\
\citet{bakos:2007}
\end{tabular} &
\hspace{-0.2cm} 
\begin{tabular}{l}
$1.163 \pm 0.069$ \\
$1.160 \pm 0.062$
\end{tabular} &
\hspace{-0.2cm} 
\begin{tabular}{l}
$1.137 \pm 0.033$ \\
$1.167 \pm 0.049$
\end{tabular} &
\hspace{-0.2cm} 
\begin{tabular}{l}
$4.392 \pm 0.020$ \\
$4.368 \pm 0.028$
\end{tabular} &
\hspace{-0.2cm} 
\begin{tabular}{c}
$0.791 \pm 0.050$ \\
$...$
\end{tabular} &
\hspace{-0.2cm} 
\begin{tabular}{l}
$1.06 \pm 0.11$ \\
$1.06 \pm 0.11$
\end{tabular} &
\hspace{-0.2cm} 
\begin{tabular}{l}
$1.252 \pm 0.043$ \\
$1.257 \pm 0.053$
\end{tabular} &
\hspace{-0.2cm} 
\begin{tabular}{l}
$16.7 \pm 1.9$ \\
$16.5 \pm 1.9$
\end{tabular} &
\hspace{-0.2cm} 
\begin{tabular}{l}
$0.504 \pm 0.067$ \\
$0.50 \pm 0.08$
\end{tabular} &
\hspace{-0.2cm} 
\begin{tabular}{l}
$0.04079 \pm 0.00080$ \\
$0.04075 \pm 0.00076$
\end{tabular}  &
\hspace{-0.2cm} 
\begin{tabular}{l}
$2.78847360\,(52)$ \\
$2.788491\,(25)$
\end{tabular} 
\\ 
\hline
HAT-P-8 & 
\begin{tabular}{l}
\citet{mancini:2013a} \\
\citet{latham:2009}
\end{tabular} &
\hspace{-0.2cm} 
\begin{tabular}{l}
$1.192 \pm 0.075$ \\
$1.28 \pm 0.04$
\end{tabular} &
\hspace{-0.2cm} 
\begin{tabular}{l}
$1.475 \pm 0.033$ \\
$1.237 \pm 0.054$
\end{tabular} &
\hspace{-0.2cm} 
\begin{tabular}{l}
$4.177 \pm 0.022$ \\
$4.15 \pm 0.03$
\end{tabular} &
\hspace{-0.2cm} 
\begin{tabular}{c}
$0.371 \pm 0.022$ \\
$...$
\end{tabular} &
\hspace{-0.2cm} 
\begin{tabular}{l}
$1.275 \pm 0.061$ \\
$1.52^{+0.18}_{-0.16}$
\end{tabular} &
\hspace{-0.2cm} 
\begin{tabular}{l}
$1.321 \pm 0.040$ \\
$1.50^{+0.08}_{-0.06}$
\end{tabular} &
\hspace{-0.2cm} 
\begin{tabular}{l}
$18.11 \pm 0.82$ \\
$16.98 \pm 1.17$
\end{tabular} &
\hspace{-0.2cm} 
\begin{tabular}{l}
$0.517 \pm 0.036$ \\
$0.568 \pm 0048$
\end{tabular} &
\hspace{-0.2cm} 
\begin{tabular}{l}
$0.04390 \pm 0.00091$ \\
$0.0487 \pm 0.0026$
\end{tabular}  &
\hspace{-0.2cm} 
\begin{tabular}{l}
$3.0763458\,(24)$ \\
$3.0763776\,(04)$
\end{tabular} 
\\ [6pt]
\hline
HAT-P-13 & 
\begin{tabular}{l}
\citet{southworth:2012a} \\
\citet{bakos:2009}
\end{tabular} &
\hspace{-0.2cm} 
\begin{tabular}{l}
$1.320 \pm 0.062$ \\
$1.219^{+0.050}_{-0.099}$
\end{tabular} &
\hspace{-0.2cm} 
\begin{tabular}{l}
$1.756 \pm 0.046$ \\
$1.559 \pm 0.082$
\end{tabular} &
\hspace{-0.2cm} 
\begin{tabular}{l}
$4.070 \pm 0.020$ \\
$4.13 \pm 0.04$
\end{tabular} &
\hspace{-0.2cm} 
\begin{tabular}{c}
$0.244 \pm 0.013$ \\
$...$
\end{tabular} &
\hspace{-0.2cm} 
\begin{tabular}{l}
$0.906 \pm 0.030$ \\
$0.853^{+0.029}_{-0.046}$
\end{tabular} &
\hspace{-0.2cm} 
\begin{tabular}{l}
$1.487 \pm 0.041$ \\
$1.281 \pm 0.079$
\end{tabular} &
\hspace{-0.2cm} 
\begin{tabular}{l}
$10.15 \pm 0.43$ \\
$12.9 \pm 1.5$
\end{tabular} &
\hspace{-0.2cm} 
\begin{tabular}{l}
$0.257 \pm 0.017$ \\
$0.375^{+0.078}_{-0.052}$
\end{tabular} &
\hspace{-0.2cm} 
\begin{tabular}{l}
$0.04383 \pm 0.00069$ \\
$0.0427^{+0.0006}_{-0.0012}$
\end{tabular}  &
\hspace{-0.2cm} 
\begin{tabular}{l}
$2.9162383\,(22)$ \\
$2.916260\,(10)$
\end{tabular} 
\\  [6pt]
\hline
HAT-P-16 & 
\begin{tabular}{l}
\citet{ciceri:2013} \\
\citet{buchhave:2010}
\end{tabular} &
\hspace{-0.2cm} 
\begin{tabular}{l}
$1.216 \pm 0.055$ \\
$1.218 \pm 0.039$
\end{tabular} &
\hspace{-0.2cm} 
\begin{tabular}{l}
$1.158 \pm 0.025$ \\
$1.237 \pm 0.054$
\end{tabular} &
\hspace{-0.2cm} 
\begin{tabular}{l}
$4.396 \pm 0.016$ \\
$4.34 \pm 0.03$
\end{tabular} &
\hspace{-0.2cm} 
\begin{tabular}{c}
$0.784 \pm 0.040$ \\
$...$
\end{tabular} &
\hspace{-0.2cm} 
\begin{tabular}{l}
$4.193 \pm 0.13$ \\
$4.193 \pm 0.094$
\end{tabular} &
\hspace{-0.2cm} 
\begin{tabular}{l}
$1.190 \pm 0.037$ \\
$1.289 \pm 0.066$
\end{tabular} &
\hspace{-0.2cm} 
\begin{tabular}{l}
$73.4 \pm 4.1$ \\
$63.1 \pm 5.8 $
\end{tabular} &
\hspace{-0.2cm} 
\begin{tabular}{l}
$2.33 \pm 0.20$ \\
$1.95 \pm 0.28$
\end{tabular} &
\hspace{-0.2cm} 
\begin{tabular}{l}
$0.04130 \pm 0.00062$ \\
$0.0413 \pm 0.0004$
\end{tabular}  &
\hspace{-0.2cm} 
\begin{tabular}{l}
$2.7759712\,(15)$ \\
$2.775960\,(03)$
\end{tabular} 
\\ 
\hline
HAT-P-23 & 
\begin{tabular}{l}
\citet{ciceri:2015a} \\
\citet{bakos:2011}
\end{tabular} &
\hspace{-0.2cm} 
\begin{tabular}{l}
$1.104 \pm 0.047$ \\
$1.13 \pm 0.04$
\end{tabular} &
\hspace{-0.2cm} 
\begin{tabular}{l}
$1.089 \pm 0.028$ \\
$1.20 \pm 0.07$
\end{tabular} &
\hspace{-0.2cm} 
\begin{tabular}{l}
$4.407 \pm 0.018$ \\
$4.33 \pm 0.06$
\end{tabular} &
\hspace{-0.2cm} 
\begin{tabular}{c}
$0.855 \pm 0.051$ \\
$...$
\end{tabular} &
\hspace{-0.2cm} 
\begin{tabular}{l}
$2.07 \pm 0.12$ \\
$2.09 \pm 0.11$
\end{tabular} &
\hspace{-0.2cm} 
\begin{tabular}{l}
$1.224 \pm 0.036$ \\
$1.368 \pm 0.090$
\end{tabular} &
\hspace{-0.2cm} 
\begin{tabular}{l}
$34.3 \pm 2.4$ \\
$27.5 \pm 3.2 $
\end{tabular} &
\hspace{-0.2cm} 
\begin{tabular}{l}
$1.057 \pm 0.097$ \\
$0.81 \pm 0.15$
\end{tabular} &
\hspace{-0.2cm} 
\begin{tabular}{l}
$0.02302 \pm 0.00032$ \\
$0.0232 \pm 0.0002$
\end{tabular}  &
\hspace{-0.2cm} 
\begin{tabular}{l}
$1.21288287\,(17)$ \\
$1.212884\,(02)$
\end{tabular} 
\\ 
\hline
HAT-P-36 & 
\begin{tabular}{l}
\citet{mancini:2015} \\
\citet{bakos:2012}
\end{tabular} &
\hspace{-0.2cm} 
\begin{tabular}{l}
$1.030 \pm 0.042$ \\
$1.022 \pm 0.049$
\end{tabular} &
\hspace{-0.2cm} 
\begin{tabular}{l}
$1.041 \pm 0.016$ \\
$1.096 \pm 0.056$
\end{tabular} &
\hspace{-0.2cm} 
\begin{tabular}{l}
$4.416 \pm 0.011$ \\
$4.37 \pm 0.04$
\end{tabular} &
\hspace{-0.2cm} 
\begin{tabular}{c}
$0.913 \pm 0.027$ \\
$...$
\end{tabular} &
\hspace{-0.2cm} 
\begin{tabular}{l}
$1.852 \pm 0.095$ \\
$1.832 \pm 0.099$
\end{tabular} &
\hspace{-0.2cm} 
\begin{tabular}{l}
$1.304 \pm 0.025$ \\
$1.264 \pm 0.071$
\end{tabular} &
\hspace{-0.2cm} 
\begin{tabular}{l}
$27.0 \pm 1.4$ \\
$28 \pm 3 $
\end{tabular} &
\hspace{-0.2cm} 
\begin{tabular}{l}
$0.737 \pm 0.095$ \\
$0.84 \pm 0.14$
\end{tabular} &
\hspace{-0.2cm} 
\begin{tabular}{l}
$0.02388 \pm 0.00032$ \\
$0.0238 \pm 0.0004$
\end{tabular}  &
\hspace{-0.2cm} 
\begin{tabular}{l}
$1.32734683\,(48)$ \\
$1.327347\,(03)$
\end{tabular} 
\\ 
\hline
Qatar-1 & 
\begin{tabular}{l}
\citet{mislis:2015} \\
\citet{alsubai:2011}
\end{tabular} &
\hspace{-0.2cm} 
\begin{tabular}{l}
$0.818 \pm 0.069$ \\
$0.85 \pm 0.03$
\end{tabular} &
 \hspace{-0.2cm}
\begin{tabular}{l}
$0.796 \pm 0.023$ \\
$0.823 \pm 0.025$
\end{tabular} &
\hspace{-0.2cm} 
\begin{tabular}{l}
$4.549 \pm 0.014$ \\
$4.536 \pm 0.024$
\end{tabular} &
\hspace{-0.2cm} 
\begin{tabular}{l}
$1.621 \pm 0.046$ \\
$1.52 \pm 0.12$
\end{tabular} &
\hspace{-0.2cm} 
\begin{tabular}{l}
$1.293 \pm 0.075$ \\
$1.090^{+0.084}_{-0.081}$
\end{tabular} &
\hspace{-0.2cm} 
\begin{tabular}{l}
$1.142 \pm 0.035$ \\
$1.164 \pm 0.045$
\end{tabular} &
\hspace{-0.2cm} 
\begin{tabular}{l}
$24.56 \pm 0.70$ \\
$18.41^{+1.86}_{-1.91} $
\end{tabular} &
\hspace{-0.2cm} 
\begin{tabular}{l}
$0.811 \pm 0.040$ \\
$0.690^{+0.098}_{-0.084}$
\end{tabular} &
\hspace{-0.2cm} 
\begin{tabular}{l}
$0.02313 \pm 0.00065$ \\
$0.02343^{+0.00026}_{-0.000 25}$
\end{tabular}  &
\hspace{-0.2cm} 
\begin{tabular}{l}
$1.4200259\,(28)$ \\
$1.420033\,(16)$
\end{tabular} 
\\ [6pt]
\hline
Qatar-2 & 
\begin{tabular}{l}
\citet{mancini:2014c} \\
\citet{bryan:2012}
\end{tabular} &
\hspace{-0.2cm} 
\begin{tabular}{l}
$0.743 \pm 0.021$ \\
$0.740 \pm 0.037$
\end{tabular} &
 \hspace{-0.2cm}
\begin{tabular}{l}
$0.776 \pm 0.008$ \\
$0.713 \pm 0.018$
\end{tabular} &
\hspace{-0.2cm} 
\begin{tabular}{l}
$4.530 \pm 0.005$ \\
$4.601 \pm 0.018$
\end{tabular} &
\hspace{-0.2cm} 
\begin{tabular}{l}
$1.591 \pm 0.016$ \\
$2.05 \pm 0.12$
\end{tabular} &
\hspace{-0.2cm} 
\begin{tabular}{l}
$2.494 \pm 0.055$ \\
$2.487 \pm 0.086$
\end{tabular} &
\hspace{-0.2cm} 
\begin{tabular}{l}
$1.254 \pm 0.013$ \\
$1.144 \pm 0.035$
\end{tabular} &
\hspace{-0.2cm} 
\begin{tabular}{l}
$39.34 \pm 0.52$ \\
$43.5 \pm 2.2$
\end{tabular} &
\hspace{-0.2cm} 
\begin{tabular}{l}
$1.183 \pm 0.022$ \\
$1.66 \pm 0.13$
\end{tabular} &
\hspace{-0.2cm} 
\begin{tabular}{l}
$0.02153 \pm 0.00020$ \\
$0.02149 \pm 0.00036$
\end{tabular}  &
\hspace{-0.2cm} 
\begin{tabular}{l}
$1.33711647\,(26)$ \\
$1.3371182\,(37)$
\end{tabular} 
\\
\hline
TrES-5 & 
\begin{tabular}{l}
\citet{mislis:2015} \\
\citet{mandushev:2011}
\end{tabular} &
\hspace{-0.2cm} 
\begin{tabular}{l}
$0.901 \pm 0.030$ \\
$0.893 \pm 0.024$
\end{tabular} &
\hspace{-0.2cm}
\begin{tabular}{l}
$0.868 \pm 0.013$ \\
$0.866 \pm 0.013$
\end{tabular} &
\hspace{-0.2cm} 
\begin{tabular}{l}
$4.517 \pm 0.012$ \\
$4.513 \pm 0.013$
\end{tabular} &
\hspace{-0.2cm} 
\begin{tabular}{c}
$1.381\pm 0.051$ \\
$...$
\end{tabular} &
\hspace{-0.2cm} 
\begin{tabular}{l}
$1.790 \pm 0.068$ \\
$1.778 \pm 0.063$
\end{tabular} &
\hspace{-0.2cm} 
\begin{tabular}{l}
$1.194 \pm 0.015$ \\
$1.209 \pm 0.021$
\end{tabular} &
\hspace{-0.2cm} 
\begin{tabular}{c}
$31.1 \pm 1.0$ \\
$... $
\end{tabular} &
\hspace{-0.2cm} 
\begin{tabular}{l}
$0.983 \pm 0.039$ \\
$1.25 \pm 0.08$
\end{tabular} &
\hspace{-0.2cm} 
\begin{tabular}{l}
$0.02459 \pm 0.00074$ \\
$0.02446 \pm 0.00068$
\end{tabular}  &
\hspace{-0.2cm} 
\begin{tabular}{l}
$1.4822469\,(61)$ \\
$1.4822446\,(07)$
\end{tabular} 
\\
\hline
WASP-2 & 
\begin{tabular}{l}
\citet{southworth:2010} \\
\citet{torres:2008}
\end{tabular} &
\hspace{-0.2cm} 
\begin{tabular}{l}
$0.803 \pm 0.060$ \\
$0.89^{+0.12}_{-0.12}$
\end{tabular} &
 \hspace{-0.2cm}
\begin{tabular}{l}
$0.807 \pm 0.022$ \\
$0.840^{+0.062}_{-0.065}$
\end{tabular} &
\hspace{-0.2cm} 
\begin{tabular}{l}
$4.529 \pm 0.018$ \\
$4.537^{+0.035}_{-0.046}$
\end{tabular} &
\hspace{-0.2cm} 
\begin{tabular}{l}
$1.527 \pm 0.067$ \\
$1.45^{+0.19}_{-0.11}$
\end{tabular} &
\hspace{-0.2cm} 
\begin{tabular}{l}
$0.847 \pm 0.045$ \\
$0.915^{+0.090}_{-0.093}$
\end{tabular} &
\hspace{-0.2cm} 
\begin{tabular}{l}
$1.043 \pm 0.033$ \\
$1.071^{+0.080}_{-0.083}$
\end{tabular} &
\hspace{-0.2cm} 
\begin{tabular}{l}
$19.32 \pm0.80$ \\
$19.4^{+1.8}_{-1.4}$
\end{tabular} &
\hspace{-0.2cm} 
\begin{tabular}{l}
$0.747 \pm 0.048$ \\
$0.74^{+0.22}_{-0.16}$
\end{tabular} &
\hspace{-0.2cm} 
\begin{tabular}{l}
$0.03033 \pm 0.00074$ \\
$0.03138^{+0.001 30}_{-0.001 54}$
\end{tabular}  &
\hspace{-0.2cm} 
\begin{tabular}{l}
$2.15222144\,(39)$ \\
$2.152226\,(04)$
\end{tabular} 
\\ [6pt]
\hline \\ [-5pt]
WASP-4 & 
\begin{tabular}{l}
\citet{southworth:2009b} \\
\citet{wilson:2008}
\end{tabular} &
\hspace{-0.2cm} 
\begin{tabular}{l}
$0.940^{+0.073}_{-0.069}$ \\
$0.8997^{+0.077}_{-0.072}$
\end{tabular} &
 \hspace{-0.2cm}
\begin{tabular}{l}
$0.914^{+0.024}_{-0.023}$\\
$0.937^{+0.04}_{-0.03}$
\end{tabular} &
\hspace{-0.2cm} 
\begin{tabular}{l}
$4.490^{+0.013}_{-0.012}$ \\
$4.450^{+0.016}_{-0.029}$
\end{tabular} &
\hspace{-0.2cm} 
\begin{tabular}{l}
$1.233^{+0.020}_{-0.022}$ \\
$1.094^{+0.038}_{-0.085}$
\end{tabular} &
\hspace{-0.2cm} 
\begin{tabular}{l}
$1.289^{+0.098}_{-0.073}$ \\
$1.215^{+0.078}_{-0.079}$
\end{tabular} &
\hspace{-0.2cm} 
\begin{tabular}{l}
$1.371^{+0.038}_{-0.035}$ \\
$1.416^{+0.068}_{-0.043}$
\end{tabular} &
\hspace{-0.2cm} 
\begin{tabular}{l}
$17.03^{+0.97}_{-0.54}$ \\
$13.87^{+0.75}_{-1.04}$
\end{tabular} &
\hspace{-0.2cm} 
\begin{tabular}{l}
$0.500^{+0.032}_{-0.022}$ \\
$0.420^{+0.032}_{-0.044}$
\end{tabular} &
\hspace{-0.2cm} 
\begin{tabular}{l}
$0.02255 \pm 0.00115$ \\
$0.02340 \pm 0.00060$
\end{tabular}  &
\hspace{-0.2cm} 
\begin{tabular}{l}
$1.33823150\,(61)$ \\
$2.152226\,(04)$
\end{tabular} 
\\ [6pt]
\hline
WASP-5 & 
\begin{tabular}{l}
\citet{southworth:2009a} \\
\citet{anderson:2008}
\end{tabular} &
\hspace{-0.2cm} 
\begin{tabular}{l}
$1.021 \pm 0.062$ \\
$0.972^{+0.099}_{-0.079}$
\end{tabular} &
 \hspace{-0.2cm}
\begin{tabular}{l}
$1.084 \pm 0.041$ \\
$1.026^{+0.073}_{-0.044}$
\end{tabular} &
\hspace{-0.2cm} 
\begin{tabular}{l}
$4.377 \pm 0.030$ \\
$4.403^{+0.039}_{-0.048}$
\end{tabular} &
\hspace{-0.2cm} 
\begin{tabular}{c}
$0.803 \pm 0.080$ \\
$...$
\end{tabular} &
\hspace{-0.2cm} 
\begin{tabular}{l}
$1.637 \pm 0.082$ \\
$1.58^{+0.13}_{-0.08}$
\end{tabular} &
\hspace{-0.2cm} 
\begin{tabular}{l}
$1.171 \pm 0.057$ \\
$1.090^{+0.094}_{-0.058}$
\end{tabular} &
\hspace{-0.2cm} 
\begin{tabular}{l}
$29.6 \pm 2.8$ \\
$30.5^{+3.2}_{-4.1}$
\end{tabular} &
\hspace{-0.2cm} 
\begin{tabular}{l}
$1.02 \pm 0.14$ \\
$1.22^{+0.19}_{-0.24}$
\end{tabular} &
\hspace{-0.2cm} 
\begin{tabular}{l}
$0.02683 \pm 0.00116$ \\
$0.0267^{+0.0012}_{-0.0008}$
\end{tabular}  &
\hspace{-0.2cm} 
\begin{tabular}{l}
$1.6284246\,(13)$ \\
$1.6284296^{(+48)}_{(-37)}$
\end{tabular} 
\\ [6pt]
\hline
WASP-6 & 
\begin{tabular}{l}
\citet{tregloanreed:2015} \\
\citet{gillon:2009}
\end{tabular} &
\hspace{-0.2cm} 
\begin{tabular}{l}
$0.836 \pm 0.067$ \\
$0.880^{+0.050}_{-0.080}$
\end{tabular} &
 \hspace{-0.2cm}
\begin{tabular}{l}
$0.864 \pm 0.025$ \\
$0.870^{+0.025}_{-0.036}$
\end{tabular} &
\hspace{-0.2cm} 
\begin{tabular}{l}
$4.487 \pm 0.017$ \\
$4.6 \pm 0.2$
\end{tabular} &
\hspace{-0.2cm} 
\begin{tabular}{l}
$1.296 \pm 0.053$ \\
$1.34^{+0.11}_{-0.10}$
\end{tabular} &
\hspace{-0.2cm} 
\begin{tabular}{l}
$0.485 \pm 0.027$ \\
$0.503^{+0.019}_{-0.038}$
\end{tabular} &
\hspace{-0.2cm} 
\begin{tabular}{l}
$1.230 \pm 0.037$ \\
$1.224^{+0.051}_{-0.052}$
\end{tabular} &
\hspace{-0.2cm} 
\begin{tabular}{l}
$7.96 \pm 0.30$ \\
$8.71\pm1.26 $
\end{tabular} &
\hspace{-0.2cm} 
\begin{tabular}{l}
$0.244 \pm 0.014$ \\
$0.27 \pm 0.05$
\end{tabular} &
\hspace{-0.2cm} 
\begin{tabular}{l}
$0.0414 \pm 0.0011$ \\
$0.0421^{+0.0008}_{-0.0013}$
\end{tabular}  &
\hspace{-0.2cm} 
\begin{tabular}{l}
$3.36100208\,(31)$ \\
$3.3610060\,(29)$
\end{tabular} 
\\ [6pt]
\hline
WASP-7 & 
\begin{tabular}{l}
\citet{southworth:2011} \\
\citet{hellier:2009}
\end{tabular} &
\hspace{-0.2cm} 
\begin{tabular}{l}
$1.276 \pm 0.065$ \\
$1.28^{+0.09}_{-0.19}$
\end{tabular} &
 \hspace{-0.2cm}
\begin{tabular}{l}
$1.432 \pm 0.092$ \\
$1.236^{+0.059}_{-0.046}$
\end{tabular} &
\hspace{-0.2cm} 
\begin{tabular}{l}
$4.232 \pm 0.047$ \\
$4.363^{+0.010}_{-0.047}$
\end{tabular} &
\hspace{-0.2cm} 
\begin{tabular}{l}
$0.434 \pm 0.074$ \\
$...$
\end{tabular} &
\hspace{-0.2cm} 
\begin{tabular}{l}
$0.96 \pm 0.13$ \\
$0.96^{+0.12}_{-0.18} $
\end{tabular} &
\hspace{-0.2cm} 
\begin{tabular}{c}
$1.330 \pm 0.093$ \\
$0.915^{+0.046}_{-0.040}$
\end{tabular} &
\hspace{-0.2cm} 
\begin{tabular}{l}
$13.4 \pm 2.6$ \\
$26.4^{+4.4}_{-4.0} $
\end{tabular} &
\hspace{-0.2cm} 
\begin{tabular}{l}
$0.41 \pm 0.10$ \\
$1.26^{+0.25}_{-0.21}$
\end{tabular} &
\hspace{-0.2cm} 
\begin{tabular}{l}
$0.0617 \pm 0.0011$ \\
$0.0618^{+0.0014}_{-0.0033}$
\end{tabular}  &
\hspace{-0.2cm} 
\begin{tabular}{l}
$4.9546416\,(35)$ \\
$ 4.954658^{(+55)}_{(-43)}$
\end{tabular} 
\\ [6pt]
\hline
WASP-11 & 
\begin{tabular}{l}
\citet{mancini:2015} \\
\citet{west:2009}
\end{tabular} &
\hspace{-0.2cm} 
\begin{tabular}{l}
$0.806 \pm 0.040$ \\
$0.77^{+0.10}_{-0.08}$
\end{tabular} &
 \hspace{-0.2cm}
\begin{tabular}{l}
$0.772 \pm 0.014$ \\
$0.74^{+0.04}_{-0.03}$
\end{tabular} &
\hspace{-0.2cm} 
\begin{tabular}{l}
$4.569 \pm 0.018$ \\
$4.45 \pm 0.20$
\end{tabular} &
\hspace{-0.2cm} 
\begin{tabular}{c}
$1.748 \pm 0.074$ \\
$...$
\end{tabular} &
\hspace{-0.2cm} 
\begin{tabular}{l}
$0.492 \pm 0.024$ \\
$0.53 \pm 0.07$
\end{tabular} &
\hspace{-0.2cm} 
\begin{tabular}{l}
$0.990 \pm 0.023$ \\
$0.91^{+0.06}_{-0.03}$
\end{tabular} &
\hspace{-0.2cm} 
\begin{tabular}{l}
$12.45 \pm 0.50$ \\
$14.45^{+1.66}_{-1.33} $
\end{tabular} &
\hspace{-0.2cm} 
\begin{tabular}{l}
$0.475 \pm 0.026$ \\
$0.69^{+0.07}_{-0.11}$
\end{tabular} &
\hspace{-0.2cm} 
\begin{tabular}{l}
$0.04375 \pm 0.00074$ \\
$0.043 \pm 0.002$
\end{tabular}  &
\hspace{-0.2cm} 
\begin{tabular}{l}
$3.72247967\,(45)$ \\
$3.722465\,(07)$
\end{tabular} 
\\ [6pt]
\hline
WASP-15 & 
\begin{tabular}{l}
\citet{southworth:2013} \\
\citet{west:2009b}
\end{tabular} &
\hspace{-0.2cm} 
\begin{tabular}{l}
$1.305 \pm 0.051$ \\
$1.18 \pm 0.12$
\end{tabular} &
 \hspace{-0.2cm}
\begin{tabular}{l}
$1.522 \pm 0.044$ \\
$1.477 \pm 0.072$
\end{tabular} &
\hspace{-0.2cm} 
\begin{tabular}{l}
$4.189 \pm 0.021$ \\
$4.169 \pm 0.033$
\end{tabular} &
\hspace{-0.2cm} 
\begin{tabular}{c}
$0.370 \pm 0.027$ \\
$0.365 \pm 0.037$
\end{tabular} &
\hspace{-0.2cm} 
\begin{tabular}{l}
$0.592 \pm 0.019$ \\
$0.542 \pm 0.050$
\end{tabular} &
\hspace{-0.2cm} 
\begin{tabular}{l}
$1.408 \pm 0.046$ \\
$1.428 \pm 0.077$
\end{tabular} &
\hspace{-0.2cm} 
\begin{tabular}{l}
$7.39 \pm 0.46$ \\
$6.08 \pm 0.62$
\end{tabular} &
\hspace{-0.2cm} 
\begin{tabular}{l}
$0.198 \pm 0.018$ \\
$0.186 \pm 0.026$
\end{tabular} &
\hspace{-0.2cm} 
\begin{tabular}{l}
$0.05165 \pm 0.00067$ \\
$0.0499 \pm 0.0018$
\end{tabular}  &
\hspace{-0.2cm} 
\begin{tabular}{l}
$3.75209748\,(81)$ \\
$3.7520656\,(28)$
\end{tabular} 
\\
\hline
WASP-16 & 
\begin{tabular}{l}
\citet{southworth:2013} \\
\citet{lister:2009}
\end{tabular} &
\hspace{-0.2cm} 
\begin{tabular}{l}
$0.980 \pm 0.054$ \\
$1.022^{+0.074}_{-0.129}$
\end{tabular} &
 \hspace{-0.2cm}
\begin{tabular}{l}
$1.087 \pm 0.042$ \\
$0.946^{+0.057}_{-0.052}$
\end{tabular} &
\hspace{-0.2cm} 
\begin{tabular}{l}
$4.357 \pm 0.022$ \\
$4.495^{+0.030}_{-0.054}$
\end{tabular} &
\hspace{-0.2cm} 
\begin{tabular}{c}
$0.762 \pm 0.056$ \\
$1.21^{+0.13}_{-0.18}$
\end{tabular} &
\hspace{-0.2cm} 
\begin{tabular}{l}
$0.832 \pm 0.038$ \\
$0.855^{+0.043}_{-0.076}$
\end{tabular} &
\hspace{-0.2cm} 
\begin{tabular}{l}
$1.218 \pm 0.040$ \\
$1.008^{+0.083}_{-0.060}$
\end{tabular} &
\hspace{-0.2cm} 
\begin{tabular}{l}
$13.92 \pm 0.71$ \\
$19.2^{+1.9}_{-2.6}$
\end{tabular} &
\hspace{-0.2cm} 
\begin{tabular}{l}
$0.431 \pm 0.033$ \\
$0.83^{+0.13}_{-0.17}$
\end{tabular} &
\hspace{-0.2cm} 
\begin{tabular}{l}
$0.04150 \pm 0.00077$ \\
$0.0421^{+0.0010}_{-0.0018}$
\end{tabular}  &
\hspace{-0.2cm} 
\begin{tabular}{l}
$3.1186068\,(12)$ \\
$3.118601^{(+15)}_{(-13)}$
\end{tabular} 
\\ [6pt]
\hline
WASP-17 & 
\begin{tabular}{l}
\citet{southworth:2012c} \\
\citet{anderson:2010}
\end{tabular} &
\hspace{-0.2cm} 
\begin{tabular}{l}
$1.286 \pm 0.079$ \\
$1.20 \pm 0.12$
\end{tabular} &
 \hspace{-0.2cm}
\begin{tabular}{l}
$1.583 \pm 0.041$ \\
$1.38^{+0.20}_{-0.18}$
\end{tabular} &
\hspace{-0.2cm} 
\begin{tabular}{l}
$4.149 \pm 0.014$ \\
$4.23 \pm 0.12$
\end{tabular} &
\hspace{-0.2cm} 
\begin{tabular}{c}
$0.324 \pm 0.012$ \\
$0.45^{+0.23}_{-0.15}$
\end{tabular} &
\hspace{-0.2cm} 
\begin{tabular}{l}
$0.477 \pm 0.033$ \\
$0.490^{+0.059}_{-0.056}$
\end{tabular} &
\hspace{-0.2cm} 
\begin{tabular}{l}
$1.932 \pm 0.053$ \\
$1.74^{+0.26}_{-0.23}$
\end{tabular} &
\hspace{-0.2cm} 
\begin{tabular}{l}
$3.16 \pm 0.20$ \\
$3.63^{+1.4}_{-0.9}$
\end{tabular} &
\hspace{-0.2cm} 
\begin{tabular}{l}
$0.0618 \pm 0.0048$ \\
$0.092^{+0.054}_{-0.032}$
\end{tabular} &
\hspace{-0.2cm} 
\begin{tabular}{l}
$0.05125 \pm 0.00103$ \\
$0.0501^{+0.0017}_{-0.0018}$
\end{tabular}  &
\hspace{-0.2cm} 
\begin{tabular}{l}
$3.735 484 5\,(19)$ \\
$3.7354417^{(+72)}_{(-73)}$
\end{tabular} 
\\ [6pt]
\hline
WASP-18 & 
\begin{tabular}{l}
\citet{southworth:2009c} \\
\citet{hellier:2009b}
\end{tabular} &
\hspace{-0.2cm} 
\begin{tabular}{l}
$1.281 \pm 0.069$ \\
$1.25 \pm 0.13$
\end{tabular} &
 \hspace{-0.2cm}
\begin{tabular}{l}
$1.230 \pm 0.047$ \\
$1.216^{+0.067}_{-0.054}$
\end{tabular} &
\hspace{-0.2cm} 
\begin{tabular}{l}
$4.366 \pm 0.026$ \\
$4.367^{+0.028}_{-0.042}$
\end{tabular} &
\hspace{-0.2cm} 
\begin{tabular}{c}
$0.689 \pm 0.062$ \\
$0.707^{+0.056}_{-0.096}$
\end{tabular} &
\hspace{-0.2cm} 
\begin{tabular}{l}
$10.43 \pm 0.38$ \\
$10.30 \pm 0.69$
\end{tabular} &
\hspace{-0.2cm} 
\begin{tabular}{l}
$1.165 \pm 0.057$ \\
$1.106^{+0.072}_{-0.054}$
\end{tabular} &
\hspace{-0.2cm} 
\begin{tabular}{l}
$191 \pm 17$ \\
$194^{+12}_{-21}$
\end{tabular} &
\hspace{-0.2cm} 
\begin{tabular}{l}
$6.60 \pm 0.90$ \\
$7.73^{+0.78}_{-1.27}$
\end{tabular} &
\hspace{-0.2cm} 
\begin{tabular}{l}
$0.02047 \pm 0.00038$ \\
$0.02026 \pm 0.00068$
\end{tabular}  &
\hspace{-0.2cm} 
\begin{tabular}{l}
$0.94145181\,(44)$ \\
$0.94145299\,(87)$
\end{tabular} 
\\ [6pt]
\hline
WASP-19 & 
\begin{tabular}{l}
\citet{mancini:2013c} \\
\citet{hebb:2010}
\end{tabular} &
\hspace{-0.2cm} 
\begin{tabular}{l}
$0.935 \pm 0.041$ \\
$0.96^{+0.09}_{-0.10}$
\end{tabular} &
 \hspace{-0.2cm}
\begin{tabular}{l}
$1.018 \pm 0.015$ \\
$0.94^{+0.04}_{-0.04}$
\end{tabular} &
\hspace{-0.2cm} 
\begin{tabular}{l}
$4.3932 \pm 0.0067$ \\
$4.47^{+0.03}_{-0.03}$
\end{tabular} &
\hspace{-0.2cm} 
\begin{tabular}{c}
$0.8853 \pm 0.0060$ \\
$1.13^{+0.09}_{-0.09}$
\end{tabular} &
\hspace{-0.2cm} 
\begin{tabular}{l}
$1.139 \pm 0.036$ \\
$1.15^{+0.08}_{-0.08}$
\end{tabular} &
\hspace{-0.2cm} 
\begin{tabular}{l}
$1.410 \pm 0.021$ \\
$1.31^{+0.06}_{-0.06}$
\end{tabular} &
\hspace{-0.2cm} 
\begin{tabular}{l}
$14.21 \pm 0.18$ \\
$15.5^{+1.1}_{-1.1} $
\end{tabular} &
\hspace{-0.2cm} 
\begin{tabular}{l}
$0.3800 \pm 0.0079$ \\
$0.51^{+0.06}_{-0.05}$
\end{tabular} &
\hspace{-0.2cm} 
\begin{tabular}{l}
$0.01634 \pm 0.00024$ \\
$0.0165^{+0.0005}_{-0.0006}$
\end{tabular}  &
\hspace{-0.2cm} 
\begin{tabular}{l}
$0.7888396\,(10)$ \\
$0.7888399^{(+08)}_{(-08)}$
\end{tabular} 
\\ [6pt]
\hline
WASP-21 & 
\begin{tabular}{l}
\citet{ciceri:2013} \\
\citet{bouchy:2010}
\end{tabular} &
\hspace{-0.2cm} 
\begin{tabular}{l}
$0.890 \pm 0.079$ \\
$1.01 \pm 0.03$
\end{tabular} &
\hspace{-0.2cm} 
\begin{tabular}{l}
$1.136 \pm 0.051$ \\
$1.06 \pm 0.04$
\end{tabular} &
\hspace{-0.2cm} 
\begin{tabular}{l}
$4.277 \pm 0.026$ \\
$4.39 \pm 0.03$
\end{tabular} &
\hspace{-0.2cm} 
\begin{tabular}{l}
$0.607 \pm 0.048$ \\
$0.84 \pm 0.09$
\end{tabular} &
\hspace{-0.2cm} 
\begin{tabular}{l}
$0.276 \pm 0.019$ \\
$0.300 \pm 0.011$
\end{tabular} &
\hspace{-0.2cm} 
\begin{tabular}{l}
$1.162 \pm 0.054$ \\
$1.07 \pm 0.06$
\end{tabular} &
\hspace{-0.2cm} 
\begin{tabular}{c}
$5.07 \pm 0.35$ \\
$...$
\end{tabular} &
\hspace{-0.2cm} 
\begin{tabular}{l}
$0.165 \pm 0.018$ \\
$0.24 \pm 0.05$
\end{tabular} &
\hspace{-0.2cm} 
\begin{tabular}{l}
$0.0499 \pm 0.0015$ \\
$0.052^{+0.00041}_{-0.00044}$
\end{tabular}  &
\hspace{-0.2cm} 
\begin{tabular}{l}
$4.3225186\,(30)$ \\
$4.322482^{(+19)}_{(-24)}$
\end{tabular} 
\\ [6pt]
\hline
WASP-22 &
\begin{tabular}{l}
\citet{southworth:2016} \\
\citet{Maxted+10aj}
\end{tabular} &
\hspace{-0.2cm}
\begin{tabular}{l}
$1.249^{+0.073}_{-0.030}$ \\
$1.1 \pm 0.3$
\end{tabular} &
\hspace{-0.2cm}
\begin{tabular}{l}
$1.255^{+0.030}_{-0.029}$ \\
$1.13 \pm 0.03$
\end{tabular} &
\hspace{-0.2cm}
\begin{tabular}{l}
$4.338^{+0.027}_{-0.020}$ \\
$4.37 \pm 0.02$
\end{tabular} &
\hspace{-0.2cm}
\begin{tabular}{l}
$0.632^{+0.043}_{-0.041}$ \\
$0.76 \pm 0.06$
\end{tabular} &
\hspace{-0.2cm}
\begin{tabular}{l}
$0.617^{+0.028}_{-0.017}$ \\
$0.56 \pm 0.02$
\end{tabular} &
\hspace{-0.2cm}
\begin{tabular}{l}
$1.199^{+0.046}_{-0.027}$ \\
$1.12 \pm 0.04$
\end{tabular} &
\hspace{-0.2cm}
\begin{tabular}{c}
$10.63^{+0.53}_{-0.71}$ \\
$10.0 \pm 0.7$
\end{tabular} &
\hspace{-0.2cm}
\begin{tabular}{l}
$0.334^{+0.024}_{-0.033}$ \\
$0.40 \pm 0.04$
\end{tabular} &
\hspace{-0.2cm}
\begin{tabular}{l}
$0.0489^{+0.0010}_{-0.0004}$ \\
$0.0468 \pm 0.0004$
\end{tabular}  &
\hspace{-0.2cm}
\begin{tabular}{l}
$3.53273064 (70)$ \\
$3.53269 (4)$
\end{tabular}
\\ [6pt]
\hline
WASP-24 & 
\begin{tabular}{l}
\citet{southworth:2014} \\
\citet{street:2010}
\end{tabular} &
\hspace{-0.2cm} 
\begin{tabular}{l}
$1.168 \pm 0.075$ \\
$1.184 \pm 0.027$
\end{tabular} &
 \hspace{-0.2cm}
\begin{tabular}{l}
$1.317 \pm 0.041$ \\
$1.331 \pm 0.032$
\end{tabular} &
\hspace{-0.2cm} 
\begin{tabular}{l}
$4.267 \pm 0.022$ \\
$4.263 \pm 0.017$
\end{tabular} &
\hspace{-0.2cm} 
\begin{tabular}{l}
$0.512 \pm 0.034$ \\
$0.502 \pm 0.030$
\end{tabular} &
\hspace{-0.2cm} 
\begin{tabular}{l}
$1.109 \pm 0.054$ \\
$1.071^{+0.036}_{-0.038}$
\end{tabular} &
\hspace{-0.2cm} 
\begin{tabular}{l}
$1.303 \pm 0.047$ \\
$1.3^{+0.039}_{-0.038}$
\end{tabular} &
\hspace{-0.2cm} 
\begin{tabular}{l}
$16.19 \pm 0.99$ \\
$14.45^{+0.87}_{-0.83} $
\end{tabular} &
\hspace{-0.2cm} 
\begin{tabular}{l}
$0.469 \pm 0.043$ \\
$0.487^{+0.043}_{-0.040}$
\end{tabular} &
\hspace{-0.2cm} 
\begin{tabular}{l}
$0.03635 \pm 0.00079$ \\
$0.03651 \pm 0.00028$
\end{tabular}  &
\hspace{-0.2cm} 
\begin{tabular}{l}
$2.3412217\,(8)$ \\
$2.34121242\,(2)$
\end{tabular} 
\\ [6pt]
\hline
WASP-25 & 
\begin{tabular}{l}
\citet{southworth:2014} \\
\citet{enoch:2010}
\end{tabular} &
\hspace{-0.2cm} 
\begin{tabular}{l}
$1.053 \pm 0.038$ \\
$1.00 \pm 0.03$
\end{tabular} &
 \hspace{-0.2cm}
\begin{tabular}{l}
$0.924 \pm 0.018$ \\
$0.92 \pm 0.04$
\end{tabular} &
\hspace{-0.2cm} 
\begin{tabular}{l}
$4.530 \pm 0.015$ \\
$4.51 \pm 0.03$
\end{tabular} &
\hspace{-0.2cm} 
\begin{tabular}{l}
$1.336 \pm 0.063$ \\
$1.29 \pm 0.10$
\end{tabular} &
\hspace{-0.2cm} 
\begin{tabular}{l}
$0.598 \pm 0.046$ \\
$0.58 \pm 0.04$
\end{tabular} &
\hspace{-0.2cm} 
\begin{tabular}{l}
$1.247 \pm 0.032$ \\
$1.22^{+0.06}_{-0.05}$
\end{tabular} &
\hspace{-0.2cm} 
\begin{tabular}{l}
$9.54 \pm 0.80$ \\
$8.91 \pm 0.82$
\end{tabular} &
\hspace{-0.2cm} 
\begin{tabular}{l}
$0.288 \pm 0.028$ \\
$0.317^{+0.036}_{-0.031}$
\end{tabular} &
\hspace{-0.2cm} 
\begin{tabular}{l}
$0.04819 \pm 0.00058$ \\
$0.0473 \pm 0.0004$
\end{tabular}  &
\hspace{-0.2cm} 
\begin{tabular}{l}
$3.7648327\,(9)$ \\
$3.764825\,(5)$
\end{tabular} 
\\
\hline
WASP-26 & 
\begin{tabular}{l}
\citet{southworth:2014} \\
\citet{smalley:2010}
\end{tabular} &
\hspace{-0.2cm} 
\begin{tabular}{l}
$1.095 \pm 0.046$ \\
$1.12 \pm 0.03$
\end{tabular} &
 \hspace{-0.2cm}
\begin{tabular}{l}
$1.284 \pm 0.036$ \\
$1.34 \pm 0.06$
\end{tabular} &
\hspace{-0.2cm} 
\begin{tabular}{l}
$4.260 \pm 0.022$ \\
$4.3 \pm 0.2$
\end{tabular} &
\hspace{-0.2cm} 
\begin{tabular}{l}
$0.517 \pm 0.037$ \\
$0.47 \pm 0.06$
\end{tabular} &
\hspace{-0.2cm} 
\begin{tabular}{l}
$1.020 \pm 0.032$ \\
$1.02 \pm 0.03$
\end{tabular} &
\hspace{-0.2cm} 
\begin{tabular}{l}
$1.216 \pm 0.047$ \\
$1.32 \pm 0.08$
\end{tabular} &
\hspace{-0.2cm} 
\begin{tabular}{l}
$17.1 \pm 1.3$ \\
$13.18 \pm 1.52$
\end{tabular} &
\hspace{-0.2cm} 
\begin{tabular}{l}
$0.530 \pm 0.060$ \\
$0.44 \pm 0.08$
\end{tabular} &
\hspace{-0.2cm} 
\begin{tabular}{l}
$0.03966 \pm 0.00056$ \\
$0.0400 \pm 0.0003$
\end{tabular}  &
\hspace{-0.2cm} 
\begin{tabular}{l}
$2.7565972\,(19)$ \\
$2.75660\,(01)$
\end{tabular} 
\\
\hline
WASP-36 & 
\begin{tabular}{l}
\citet{mancini:2016} \\
\citet{smith:2012} 
\end{tabular} &
\hspace{-0.2cm} 
\begin{tabular}{l}
$1.081 \pm 0.034$ \\
$1.040 \pm 0.031$
\end{tabular} &
 \hspace{-0.2cm}
\begin{tabular}{l}
$0.985 \pm 0.014$ \\
$0.951 \pm 0.018$
\end{tabular} &
\hspace{-0.2cm} 
\begin{tabular}{l}
$4.486 \pm 0.0095$ \\
$4.499 \pm 0.012$
\end{tabular} &
\hspace{-0.2cm} 
\begin{tabular}{l}
$1.132 \pm 0.032$ \\
$1.211 \pm 0.050$
\end{tabular} &
\hspace{-0.2cm} 
\begin{tabular}{l}
$2.361 \pm 0.070$ \\
$2.303 \pm0.068$
\end{tabular} &
\hspace{-0.2cm} 
\begin{tabular}{l}
$1.327 \pm 0.021$ \\
$1.281 \pm 0.029$
\end{tabular} &
\hspace{-0.2cm} 
\begin{tabular}{l}
$33.2 \pm 1.1$ \\
$32.1 \pm1.3$
\end{tabular} &
\hspace{-0.2cm} 
\begin{tabular}{l}
$0.945 \pm 0.042$ \\
$1.096 \pm 0.067$
\end{tabular} &
\hspace{-0.2cm} 
\begin{tabular}{l}
$0.02677 \pm 0.00028$ \\
$0.02643 \pm 0.00026$
\end{tabular}  &
\hspace{-0.2cm} 
\begin{tabular}{l}
$1.53736590\,(28)$ \\
$1.5373653\,(26)$
\end{tabular} 
\\
\hline
WASP-41 &
\begin{tabular}{l}
\citet{southworth:2016} \\
\citet{Maxted+11pasp}
\end{tabular} &
\hspace{-0.2cm}
\begin{tabular}{l}
$0.987 \pm 0.021$ \\
$0.94 \pm 0.03$
\end{tabular} &
\hspace{-0.2cm}
\begin{tabular}{l}
$0.886 \pm 0.009$ \\
$0.91 \pm 0.05$
\end{tabular} &
\hspace{-0.2cm}
\begin{tabular}{l}
$4.538 \pm 0.008$ \\
$4.4 \pm 0.2$
\end{tabular} &
\hspace{-0.2cm}
\begin{tabular}{l}
$1.420 \pm 0.034$ \\
$1.27 \pm 0.14$
\end{tabular} &
\hspace{-0.2cm}
\begin{tabular}{l}
$0.977 \pm 0.020$ \\
$0.92 \pm 0.07$
\end{tabular} &
\hspace{-0.2cm}
\begin{tabular}{l}
$1.178 \pm 0.015$ \\
$1.21 \pm 0.07$
\end{tabular} &
\hspace{-0.2cm}
\begin{tabular}{c}
$17.45 \pm 0.46$ \\
$14.5 \pm 1.3$
\end{tabular} &
\hspace{-0.2cm}
\begin{tabular}{l}
$0.558 \pm 0.020$ \\
$0.49 \pm 0.08$
\end{tabular} &
\hspace{-0.2cm}
\begin{tabular}{l}
$0.0410 \pm 0.0003$ \\
$0.0403 \pm 0.0005$
\end{tabular}  &
\hspace{-0.2cm}
\begin{tabular}{l}
$3.05240154 (41)$ \\
$3.052394 (4)$
\end{tabular}
\\ [6pt]
\hline
WASP-42 &
\begin{tabular}{l}
\citet{southworth:2016} \\
\citet{Lendl+12aa}
\end{tabular} &
\hspace{-0.2cm}
\begin{tabular}{l}
$0.951 \pm 0.037$ \\
$0.884^{+0.086}_{-0.080}$
\end{tabular} &
\hspace{-0.2cm}
\begin{tabular}{l}
$0.892 \pm 0.021$ \\
$0.863^{+0.041}_{-0.034}$
\end{tabular} &
\hspace{-0.2cm}
\begin{tabular}{l}
$4.515 \pm 0.022$ \\
$4.5 \pm 0.1$
\end{tabular} &
\hspace{-0.2cm}
\begin{tabular}{l}
$1.338 \pm 0.092$ \\
$1.37 \pm 0.14$
\end{tabular} &
\hspace{-0.2cm}
\begin{tabular}{l}
$0.527 \pm 0.020$ \\
$0.500 \pm 0.035$
\end{tabular} &
\hspace{-0.2cm}
\begin{tabular}{l}
$1.122 \pm 0.033$ \\
$1.080 \pm 0.057$
\end{tabular} &
\hspace{-0.2cm}
\begin{tabular}{c}
$10.38 \pm 0.61$ \\
$...$
\end{tabular} &
\hspace{-0.2cm}
\begin{tabular}{l}
$0.349 \pm 0.029$ \\
$0.397^{+0.054}_{-0.047}$
\end{tabular} &
\hspace{-0.2cm}
\begin{tabular}{l}
$0.0561 \pm 0.0007$ \\
$0.0548 \pm 0.0017$
\end{tabular}  &
\hspace{-0.2cm}
\begin{tabular}{l}
$4.9816819 (11)$ \\
$4.9816872 (73)$
\end{tabular}
\\ [6pt]
\hline
WASP-44 & 
\begin{tabular}{l}
\citet{mancini:2013b} \\
\citet{anderson:2012}
\end{tabular} &
\hspace{-0.2cm} 
\begin{tabular}{l}
$0.917 \pm 0.092$ \\
$0.951 \pm 0.034$
\end{tabular} &
\hspace{-0.2cm} 
\begin{tabular}{l}
$0.865 \pm 0.030$ \\
$0.927^{+0.068}_{-0.074}$
\end{tabular} &
\hspace{-0.2cm} 
\begin{tabular}{l}
$4.526 \pm 0.020$ \\
$4.481^{+0.068}_{-0.057}$
\end{tabular} &
\hspace{-0.2cm} 
\begin{tabular}{l}
$1.414 \pm 0.058$ \\
$1.19^{+0.32}_{-0.22}$
\end{tabular} &
\hspace{-0.2cm} 
\begin{tabular}{l}
$0.869 \pm 0.082$ \\
$0.889 \pm 0.062$
\end{tabular} &
\hspace{-0.2cm} 
\begin{tabular}{l}
$1.002 \pm 0.038$ \\
$1.14 \pm 0.11$
\end{tabular} &
\hspace{-0.2cm} 
\begin{tabular}{l}
$21.5 \pm 1.6$ \\
$15.7^{+3.4}_{-3.0}$
\end{tabular} &
\hspace{-0.2cm} 
\begin{tabular}{l}
$0.808 \pm 0.074$ \\
$0.61^{+0.23}_{-0.15}$
\end{tabular} &
\hspace{-0.2cm} 
\begin{tabular}{l}
$0.03445 \pm 0.00112$ \\
$0.03473 \pm 0.00041$
\end{tabular}  &
\hspace{-0.2cm} 
\begin{tabular}{l}
$2.4238133\,(23)$ \\
$2.4238039\,(87)$
\end{tabular} 
\\ [6pt]
\hline
WASP-45 & 
\begin{tabular}{l}
\citet{ciceri:2015b} \\
\citet{anderson:2012}
\end{tabular} &
\hspace{-0.2cm} 
\begin{tabular}{l}
$0.904 \pm 0.067$ \\
$0.909 \pm 0.060$
\end{tabular} &
\hspace{-0.2cm} 
\begin{tabular}{l}
$0.917 \pm 0.024$ \\
$0.945^{+0.87}_{-0.71}$
\end{tabular} &
\hspace{-0.2cm} 
\begin{tabular}{l}
$4.470 \pm 0.014$ \\
$4.445^{+0.065}_{-0.075}$
\end{tabular} &
\hspace{-0.2cm} 
\begin{tabular}{l}
$1.174 \pm 0.047$ \\
$1.08^{+0.27}_{-0.24}$
\end{tabular} &
\hspace{-0.2cm} 
\begin{tabular}{l}
$1.002 \pm 0.062$ \\
$1.007 \pm 0.053$
\end{tabular} &
\hspace{-0.2cm} 
\begin{tabular}{l}
$0.992 \pm 0.038$ \\
$1.16^{+0.28}_{-0.14}$
\end{tabular} &
\hspace{-0.2cm} 
\begin{tabular}{l}
$25.2 \pm 1.3$ \\
$17.0^{+4.9}_{-6.0}$
\end{tabular} &
\hspace{-0.2cm} 
\begin{tabular}{l}
$0.959 \pm 0.077$ \\
$0.64 \pm 0.30$
\end{tabular} &
\hspace{-0.2cm} 
\begin{tabular}{l}
$0.0405 \pm 0.0010$ \\
$0.04054 \pm 0.00090$
\end{tabular}  &
\hspace{-0.2cm} 
\begin{tabular}{l}
$3.12609601\,(49)$ \\
$3.1260876\,(35)$
\end{tabular} 
\\ [6pt]
\hline
WASP-46 & 
\begin{tabular}{l}
\citet{ciceri:2015b} \\
\citet{anderson:2012}
\end{tabular} &
\hspace{-0.2cm} 
\begin{tabular}{l}
$0.828 \pm 0.076$ \\
$0.956 \pm0.034$
\end{tabular} &
\hspace{-0.2cm} 
\begin{tabular}{l}
$0.858 \pm 0.027$ \\
$0.917 \pm 0.028$
\end{tabular} &
\hspace{-0.2cm} 
\begin{tabular}{l}
$4.489 \pm 0.014$ \\
$4.493 \pm 0.023$
\end{tabular} &
\hspace{-0.2cm} 
\begin{tabular}{l}
$1.310 \pm 0.025$ \\
$1.24 \pm0.10$
\end{tabular} &
\hspace{-0.2cm} 
\begin{tabular}{l}
$1.91 \pm 0.13$ \\
$2.101 \pm 0.073$
\end{tabular} &
\hspace{-0.2cm} 
\begin{tabular}{l}
$1.174 \pm 0.037$ \\
$1.310 \pm 0.051$
\end{tabular} &
\hspace{-0.2cm} 
\begin{tabular}{l}
$34.3 \pm 1.1$ \\
$28.0^{+2.2}_{-2.0}$
\end{tabular} &
\hspace{-0.2cm} 
\begin{tabular}{l}
$1.103 \pm 0.052$ \\
$0.94 \pm 0.11$
\end{tabular} &
\hspace{-0.2cm} 
\begin{tabular}{l}
$0.02335 \pm 0.00072$ \\
$0.02448 \pm 0.00028$
\end{tabular}  &
\hspace{-0.2cm} 
\begin{tabular}{l}
$1.43036763\,(93)$ \\
$1.4303700\,(23)$
\end{tabular} 
\\ [6pt]
\hline
WASP-48 & 
\begin{tabular}{l}
\citet{ciceri:2015a} \\
\citet{enoch:2011}
\end{tabular} &
\hspace{-0.2cm} 
\begin{tabular}{l}
$1.062 \pm 0.075$ \\
$1.19 \pm 0.05$
\end{tabular} &
\hspace{-0.2cm} 
\begin{tabular}{l}
$1.519 \pm 0.051$ \\
$1.75 \pm 0.09$
\end{tabular} &
\hspace{-0.2cm} 
\begin{tabular}{l}
$4.101 \pm 0.023$ \\
$4.03 \pm 0.04$
\end{tabular} &
\hspace{-0.2cm} 
\begin{tabular}{l}
$0.303 \pm 0.022$ \\
$0.22 \pm 0.03$
\end{tabular} &
\hspace{-0.2cm} 
\begin{tabular}{l}
$0.907 \pm 0.085$ \\
$0.98 \pm 0.09$
\end{tabular} &
\hspace{-0.2cm} 
\begin{tabular}{l}
$1.396 \pm 0.051$ \\
$1.67 \pm 0.10$
\end{tabular} &
\hspace{-0.2cm} 
\begin{tabular}{l}
$11.5 \pm 1.1$ \\
$8.1 \pm 1.1$
\end{tabular} &
\hspace{-0.2cm} 
\begin{tabular}{l}
$0.312 \pm 0.037$ \\
$0.21 \pm 0.04$
\end{tabular} &
\hspace{-0.2cm} 
\begin{tabular}{l}
$0.03320 \pm 0.00077$ \\
$0.03444 \pm 0.00043$
\end{tabular}  &
\hspace{-0.2cm} 
\begin{tabular}{l}
$2.14363544\,(58)$ \\
$2.143634\,(03)$
\end{tabular} 
\\
\hline
WASP-50 &
\begin{tabular}{l}
\citet{TregloanSouthworth13mn} \\
\citet{Gillon+11aa}
\end{tabular} &
\hspace{-0.2cm}
\begin{tabular}{l}
$0.861 \pm 0.057$ \\
$0.892^{+0.080}_{-0.074}$
\end{tabular} &
\hspace{-0.2cm}
\begin{tabular}{l}
$0.855 \pm 0.019$ \\
$0.843 \pm 0.031$
\end{tabular} &
\hspace{-0.2cm}
\begin{tabular}{l}
$4.509 \pm 0.013$ \\
$4.537 \pm 0.022$
\end{tabular} &
\hspace{-0.2cm}
\begin{tabular}{l}
$1.376 \pm 0.032$ \\
$1.48^{+0.10}_{-0.09}$
\end{tabular} &
\hspace{-0.2cm}
\begin{tabular}{l}
$1.437 \pm 0.068$ \\
$1.168^{+0.091}_{-0.086}$
\end{tabular} &
\hspace{-0.2cm}
\begin{tabular}{l}
$1.138 \pm 0.026$ \\
$1.153 \pm 0.048$
\end{tabular} &
\hspace{-0.2cm}
\begin{tabular}{l}
$27.50 \pm 0.64$ \\
$27.5 \pm 1.6$
\end{tabular} &
\hspace{-0.2cm}
\begin{tabular}{l}
$0.911 \pm 0.033$ \\
$0.958^{+0.095}_{0.082}$
\end{tabular} &
\hspace{-0.2cm}
\begin{tabular}{l}
$0.02913 \pm 0.00064$ \\
$0.02945 \pm 0.00085$
\end{tabular}  &
\hspace{-0.2cm}
\begin{tabular}{l}
$1.9550938 (13)$ \\
$1.9550959 (51)$
\end{tabular}
\\ [6pt]
\hline
WASP-55 &
\begin{tabular}{l}
\citet{southworth:2016} \\
\citet{Hellier+12mn}
\end{tabular} &
\hspace{-0.2cm}
\begin{tabular}{l}
$1.162^{+0.029}_{-0.033}$ \\
$1.01 \pm 0.04$
\end{tabular} &
\hspace{-0.2cm}
\begin{tabular}{l}
$1.102^{+0.020}_{-0.015}$ \\
$1.06^{+0.03}_{-0.02}$
\end{tabular} &
\hspace{-0.2cm}
\begin{tabular}{l}
$4.419^{+0.009}_{-0.015}$ \\
$4.39^{+0.01}_{-0.02}$
\end{tabular} &
\hspace{-0.2cm}
\begin{tabular}{l}
$0.869^{+0.026}_{-0.041}$ \\
$0.85^{+0.03}_{-0.07}$
\end{tabular} &
 \hspace{-0.2cm}
\begin{tabular}{l}
$0.627^{+0.037}_{-0.038}$ \\
$0.57 \pm 0.04$
\end{tabular} &
\hspace{-0.2cm}
\begin{tabular}{l}
$1.335^{+0.031}_{-0.020}$ \\
$1.30^{+0.05}_{-0.03}$
\end{tabular} &
\hspace{-0.2cm}
\begin{tabular}{c}
$8.73^{+0.54}_{-0.62}$ \\
$7.76 \pm 0.72$
\end{tabular} &
\hspace{-0.2cm}
\begin{tabular}{l}
$0.247^{+0.017}_{-0.021}$ \\
$0.26^{+0.02}_{-0.03}$
\end{tabular} &
\hspace{-0.2cm}
\begin{tabular}{l}
$0.0558 \pm 0.0005$ \\
$0.0433 \pm 0.0007$
\end{tabular}  &
\hspace{-0.2cm}
\begin{tabular}{l}
$4.4656291 (11)$ \\
$4.465633 (4)$
\end{tabular}
\\ [6pt]
\hline
WASP-57 & 
\begin{tabular}{l}
\citet{southworth:2015b} \\
\citet{faedi:2013}
\end{tabular} &
\hspace{-0.2cm} 
\begin{tabular}{l}
$0.886 \pm 0.067$ \\
$0.954 \pm 0.028$
\end{tabular} &
\hspace{-0.2cm} 
\begin{tabular}{l}
$0.927 \pm 0.033$ \\
$0.836^{+0.07}_{-0.16}$
\end{tabular} &
\hspace{-0.2cm} 
\begin{tabular}{l}
$4.452 \pm 0.025$ \\
$4.574^{+0.009}_{-0.012}$
\end{tabular} &
\hspace{-0.2cm} 
\begin{tabular}{l}
$1.113 \pm 0.085$ \\
$1.638^{+0.044}_{-0.063}$
\end{tabular} &
\hspace{-0.2cm} 
\begin{tabular}{l}
$0.644 \pm 0.062$ \\
$0.672^{+0.049}_{-0.046}$
\end{tabular} &
\hspace{-0.2cm} 
\begin{tabular}{l}
$1.050 \pm 0.053$ \\
$0.916^{+0.017}_{-0.014}$
\end{tabular} &
\hspace{-0.2cm} 
\begin{tabular}{l}
$14.5 \pm 1.5$ \\
$18.3^{+2.9}_{-1.3}$
\end{tabular} &
\hspace{-0.2cm} 
\begin{tabular}{l}
$0.521 \pm 0.072$ \\
$0.873^{+0.076}_{-0.071}$
\end{tabular} &
\hspace{-0.2cm} 
\begin{tabular}{l}
$0.03769 \pm 0.00088$ \\
$0.0386 \pm 0.0004$
\end{tabular}  &
\hspace{-0.2cm} 
\begin{tabular}{l}
$2.83891856\,(81)$ \\
$2.838971\,(02)$
\end{tabular} 
\\ [6pt]
\hline
WASP-67 & 
\begin{tabular}{l}
\citet{mancini:2014b} \\
\citet{faedi:2013}
\end{tabular} &
\hspace{-0.2cm} 
\begin{tabular}{l}
$0.829 \pm 0.062$ \\
$0.87 \pm 0.04$
\end{tabular} &
\hspace{-0.2cm} 
\begin{tabular}{l}
$0.817 \pm 0.022$ \\
$0.87 \pm 0.04$
\end{tabular} &
\hspace{-0.2cm} 
\begin{tabular}{l}
$4.533 \pm 0.016$ \\
$4.50 \pm 0.03$
\end{tabular} &
\hspace{-0.2cm} 
\begin{tabular}{l}
$1.522 \pm 0.049$ \\
$1.32 \pm 0.15$
\end{tabular} &
\hspace{-0.2cm} 
\begin{tabular}{l}
$0.406 \pm 0.035$ \\
$0.42 \pm 0.04$
\end{tabular} &
\hspace{-0.2cm} 
\begin{tabular}{l}
$1.091 \pm 0.046$ \\
$1.4^{+0.3}_{-0.2}$
\end{tabular} &
\hspace{-0.2cm} 
\begin{tabular}{l}
$8.45 \pm 0.83$ \\
$5.0^{+1.2}_{-2.3}$
\end{tabular} &
\hspace{-0.2cm} 
\begin{tabular}{l}
$0.292 \pm 0.036$ \\
$0.16 \pm 0.08$
\end{tabular} &
\hspace{-0.2cm} 
\begin{tabular}{l}
$0.0510 \pm 0.0012$ \\
$0.0517 \pm 0.0008$
\end{tabular}  &
\hspace{-0.2cm} 
\begin{tabular}{l}
$4.6144109\,(27)$ \\
$4.61442\,(01)$
\end{tabular} 
\\ [6pt]
\hline
WASP-80 & 
\begin{tabular}{l}
\citet{mancini:2014a} \\
\citet{triaud:2013}
\end{tabular} &
\hspace{-0.2cm} 
\begin{tabular}{l}
$0.596 \pm 0.035$ \\
$0.57^{+0.05}_{-0.05}$
\end{tabular} &
\hspace{-0.2cm} 
\begin{tabular}{l}
$0.593 \pm 0.012$ \\
$0.571^{+0.016}_{-0.016}$
\end{tabular} &
\hspace{-0.2cm} 
\begin{tabular}{l}
$4.6678 \pm 0.0084$ \\
$4.689^{+0.012}_{-0.013}$
\end{tabular} &
\hspace{-0.2cm} 
\begin{tabular}{l}
$2.862 \pm 0.050$ \\
$3.117^{+0.021}_{-0.020}$
\end{tabular} &
\hspace{-0.2cm} 
\begin{tabular}{l}
$0.562 \pm 0.027$ \\
$0.554^{+0.030}_{-0.039}$
\end{tabular} &
\hspace{-0.2cm} 
\begin{tabular}{l}
$0.986 \pm 0.022$ \\
$0.952^{+0.026}_{-0.027}$
\end{tabular} &
\hspace{-0.2cm} 
\begin{tabular}{l}
$14.34 \pm 0.46$ \\
$15.07^{+0.45}_{-0.42}$
\end{tabular} &
\hspace{-0.2cm} 
\begin{tabular}{l}
$0.549 \pm 0.023$ \\
$0.554^{+0.030}_{-0.039}$
\end{tabular} &
\hspace{-0.2cm} 
\begin{tabular}{l}
$0.03479 \pm 0.00068$ \\
$0.0346^{+0.008}_{-0.011}$
\end{tabular}  &
\hspace{-0.2cm} 
\begin{tabular}{l}
$3.06786144\,(87)$ \\
$3.0678504^{(+23)}_{(-27)}$
\end{tabular} 
\\ [6pt]
\hline
WASP-103 & 
\begin{tabular}{l}
\citet{southworth:2015a} \\
\citet{gillon:2014}
\end{tabular} &
\hspace{-0.2cm} 
\begin{tabular}{l}
$1.204 \pm 0.091$ \\
$1.220^{+0.039}_{-0.036}$
\end{tabular} &
\hspace{-0.2cm} 
\begin{tabular}{l}
$1.419 \pm 0.040$ \\
$1.436^{+0.052}_{-0.031}$
\end{tabular} &
\hspace{-0.2cm} 
\begin{tabular}{l}
$4.215 \pm 0.014$ \\
$4.22^{+0.12}_{-0.05}$
\end{tabular} &
\hspace{-0.2cm} 
\begin{tabular}{l}
$0.421 \pm 0.013$ \\
$0.414^{+0.021}_{-0.039}$
\end{tabular} &
\hspace{-0.2cm} 
\begin{tabular}{l}
$1.47 \pm 0.11$ \\
$1.490 \pm 0.088$
\end{tabular} &
\hspace{-0.2cm} 
\begin{tabular}{l}
$1.554 \pm 0.045$ \\
$1.528^{+0.073}_{-0.047}$
\end{tabular} &
\hspace{-0.2cm} 
\begin{tabular}{l}
$15.12 \pm 0.93$ \\
$15.7 \pm 1.4$
\end{tabular} &
\hspace{-0.2cm} 
\begin{tabular}{l}
$0.367 \pm 0.027$ \\
$0.415^{+0.046}_{-0.053}$
\end{tabular} &
\hspace{-0.2cm} 
\begin{tabular}{l}
$0.01978 \pm 0.00050$ \\
$0.019 85 \pm 0.000 21$
\end{tabular}  &
\hspace{-0.2cm} 
\begin{tabular}{l}
$0.9255456\,(13)$ \\
$0.925542\,(19)$
\end{tabular} 
\\ [6pt]
\hline
\end{longtable}
}
\end{center}
\end{landscape}





\bibliographystyle{aa} 

\end{document}